\newif\ifprbclass
\IfFileExists{revtex4-2.cls}{\prbclasstrue}{\prbclassfalse}
\ifprbclass
  \documentclass[
    10pt,
    aps,
    prb,
    reprint,
    superscriptaddress,
    longbibliography,
    floatfix
  ]{revtex4-2}
\else
  \documentclass[10pt,twocolumn]{article}
  \usepackage[
    letterpaper,
    margin=0.70in,
    columnsep=0.25in
  ]{geometry}
\fi
\usepackage{silence}
\usepackage{amsmath,amssymb,mathtools}
\usepackage{braket}
\usepackage{graphicx}
\usepackage{microtype}
\usepackage{orcidlink}
\usepackage{hyperref}
\hypersetup{hidelinks}
\graphicspath{
  {main/figures/}
  {appendix/figures/}
}
\begin{document}
\title{Rescaled Mandelstam–Tamm characterization of discrete
time-crystal response in a disordered Floquet Ising chain}
\ifprbclass
  \date[]{}
\else
  \date{}
\fi
\ifprbclass
\author{Abrar Ahmed Naqash\,
\orcidlink{0000-0003-3891-4740}}
\email[Corresponding author: ]{abrar.naqash@candqrc.ca}
\affiliation{Canadian Quantum Research Center,
204-3002 32 Ave, Vernon, BC V1T 2L7, Canada}
\author{Salman Sajad Wani\,
\orcidlink{0000-0002-5262-9738}}
\email{sawa54922@hbku.edu.qa}
\affiliation{Qatar Center for Quantum Computing,
College of Science and Engineering,
Hamad Bin Khalifa University, Doha, Qatar}

\author{Saif Al-Kuwari\,
\orcidlink{0000-0002-4402-7710}}
\email{smalkuwari@hbku.edu.qa}
\affiliation{Qatar Center for Quantum Computing,
College of Science and Engineering,
Hamad Bin Khalifa University, Doha, Qatar}
\else
\author{
Salman Sajad Wani\,\orcidlink{0000-0002-5262-9738},
Abrar Ahmed Naqash\,\orcidlink{0000-0003-3891-4740}, and
Saif Al-Kuwari\,\orcidlink{0000-0002-4402-7710}
\\[0.5em]
\small
Qatar Center for Quantum Computing,
College of Science and Engineering,
\\[-0.1em]
\small
Hamad Bin Khalifa University, Doha, Qatar
\\[0.25em]
\small
Canadian Quantum Research Center,
204-3002 32 Ave, Vernon, BC V1T 2L7, Canada
\\[0.4em]
\footnotesize
sawa54922@hbku.edu.qa;
\textit{Corresponding author:}
abrar.naqash@candqrc.ca;
smalkuwari@hbku.edu.qa
}
\fi
\newcommand{\manuscriptabstract}{For pure-state unitary dynamics, the Mandelstam–Tamm (MT) lower-bound functional compares the endpoint Fubini–Study return angle with path-averaged energy dispersion. Their distinct size and temporal dependences obscure the origin of period-two MT structure in discrete time-crystal (DTC)-like dynamics and its relation to the spin response. For binary Floquet drives, we derive an exact segment-resolved MT expression without assuming commuting segment Hamiltonians and apply it to a disordered Floquet Ising chain. Absolute and uniform summability of connected covariances of local energy terms implies an $O(\sqrt{L})$ upper bound on the path-averaged energy dispersion. Endpoint data for four system sizes are consistent with this leading behavior and support the corresponding rescaling of the MT functional. At a representative point in the finite-size region with a locked spin response, odd and even rescaled MT branches remain separated throughout the $10^2$-period observation window. The return angle alternates strongly,
whereas the size-normalized path-averaged energy dispersion shows little discernible parity dependence, indicating that endpoint geometry is the main source of the branch splitting. Across the interacting parameter grid, the period-two MT component has a strong partial Spearman rank correlation with the locked spin response after controlling for pulse error and interaction strength. The rescaled MT functional characterizes the global return geometry of finite-size period-two dynamics and quantifies its association with the locked spin response.
}
\ifprbclass
\begin{abstract}
\manuscriptabstract
\end{abstract}
\maketitle
\else
\maketitle
\begin{abstract}
\manuscriptabstract
\end{abstract}
\fi
\section{Introduction}
\label{sec:introduction}

Periodic driving can generate nonequilibrium phases. Generic isolated interacting Floquet systems absorb energy under sustained driving, and local
observables approach their infinite-temperature values
\cite{DAlessioRigol2014,LazaridesDasMoessner2014}. Quenched disorder can induce Floquet many-body localization and suppress this heating \cite{Ponte2015,Khemani2016,Sierant2023}. At sufficiently high drive frequencies, a prethermal regime can delay heating for parametrically long
times \cite{ElsePrethermal2017,AbaninDeRoeckHoHuveneers2017}. Discrete time-crystalline response can consequently persist in a localized regime or throughout a long prethermal window. A discrete time crystal spontaneously breaks the discrete time-translation symmetry of the drive and exhibits a rigid subharmonic response \cite{Else2016,Khemani2016,vonKeyserlingk2016,Yao2017,Zaletel2023}. A key signature of this rigidity is locking of the subharmonic frequency over a finite range of pulse-angle perturbations \cite{Yao2017,Zaletel2023}. In disordered binary Floquet Ising chains, interactions stabilize the locked
response against pulse errors, and localization suppresses heating \cite{Khemani2016,Yao2017,Iemini2024}. Experiments with trapped ions and disordered
dipolar spins have observed related subharmonic responses \cite{Zhang2017,Choi2017}. In the localized Floquet setting, DTC order is characterized by persistent subharmonic correlations, long-range eigenstate order, and pairs of Floquet eigenstates whose eigenphases differ by $\pi$ \cite{Else2016,Khemani2016,vonKeyserlingk2016}.

For the random-field Floquet Ising model, Schmid et al.~\cite{Schmid2024} found that spectral $\pi$ pairing remains robust under longitudinal-field disorder in finite chains. Penner et al.~\cite{Penner2025} found that the typical finite-size
deviation from exact pairing decreases exponentially with system size. They further showed that the disorder-averaged temporal spin correlations are
proportional to the Fourier transform of the deviation distribution. The adjacent-gap ratio probes quasienergy level repulsion and tracks the finite-size crossover between localized-like and thermal-like spectra \cite{OganesyanHuse2007,Atas2013,Ponte2015}. Using a geometric Floquet construction, Schindler and Bukov~\cite{SchindlerBukov2025} traced the $\pi$ eigenphase separation between even- and odd-parity partners to the difference of their geometric
phases. Nonanalyticities in the return probability \cite{Heyl2017,Suman2024}
have also been used to diagnose dynamical quantum phase transitions following a quench in a DTC \cite{KosiorSacha2018}.

The Mandelstam--Tamm (MT) relation, originally derived for a time-independent Hamiltonian \cite{MandelstamTamm1945}, {admits a geometric formulation in which the Fubini--Study distance between the initial and final state rays is bounded above by the length of the state trajectory in projective Hilbert space}. For pure states evolving under a time-dependent Hamiltonian, {this formulation yields a lower bound on the elapsed time in terms of} the endpoint Fubini--Study return angle and the path-averaged energy dispersion \cite{AnandanAharonov1990,DeffnerLutz2013}. {The return fidelity determines the endpoint angle} \cite{Giovannetti2003,Fogarty2020}. {The time-integrated energy dispersion determines the Fubini--Study length of the actual state trajectory. The MT relation therefore incorporates this path length together with the endpoint angle supplied by return fidelity. We examine the respective roles of these quantities in the finite-time period-two structure of the stroboscopic MT sequence and in its variation across the drive-parameter grid.}

After accounting for the many-body size dependence, we investigate what produces the period-two structure of the rescaled stroboscopic MT sequence for an initial state and how this structure relates to the locked spin response. At finite size and over the chosen observation window, we classify a parameter point as DTC-like when it exhibits a locked period-two spin response and localized-like adjacent-gap statistics. For a piecewise-constant binary drive in which $H_1$ and $H_2$ act successively for arbitrary positive durations, we derive an exact segment-resolved expression for the pure-state MT lower-bound functional. The time-integrated energy dispersion becomes a duration-weighted sum of the dispersions evaluated in the states entering each segment. The expression also applies when $[H_1,H_2]\neq0$. Suppose that each segment Hamiltonian is composed of $O(L)$ uniformly bounded local terms. If their symmetrized connected covariances are absolutely summable with a bound uniform in system size, disorder realization, drive segment, and time throughout the observation window, then the energy variance is $O(L)$. The instantaneous and path-averaged energy dispersions are consequently $O(\sqrt L)$.

{At the fixed endpoint used for the finite-size comparison, the disorder mean of the path-averaged energy dispersion is consistent with leading $\sqrt L$ scaling across four sizes. This supports the $\sqrt L$ rescaling used for the dimensionless MT diagnostic. In the time-resolved finite-size data, the rescaled MT sequence shows period-two structure at parameter points where the spin response is locked. At a representative DTC-like point, the MT values at odd and even stroboscopic indices form separated branches throughout the observation window. Most of this separation arises from strong endpoint-angle alternation; the size-normalized energy dispersion, averaged from the initial time to each stroboscopic endpoint, shows little discernible odd--even modulation. At the fixed endpoint, we identify a subset in which the return angles lie close to their maximum, $\pi/2$. Across this subset, the normalized path-averaged energy dispersion governs most of the variation in the rescaled MT diagnostic. The diagnostic consequently remains sensitive to differences in accumulated Fubini--Study path length among evolutions with similar endpoint fidelities.}

Across the interacting finite-size parameter grid, the partial Spearman rank correlation between the locked MT component and the locked spin response remains strong after jointly controlling for pulse error and interaction strength. Both observables are computed from the same trajectories using the same finite-window projection onto the period-two Fourier component. Their strong correlation links the finite-time subharmonic components of the MT sequence and the spin response. On the nested interacting grid used for the spectral analysis, the locked MT component has a much weaker partial Spearman rank correlation with the disorder-averaged adjacent-gap ratio after the same joint control. In this finite-size, finite-time setting, the rescaled MT sequence provides a state-dependent geometric characterization of the period-two dynamics tracked by the locked spin response.

\section{Mandelstam–Tamm bound for Floquet evolution}
\label{sec:floquet_mt}
For each finite system size, let $H(t)$ be a Hermitian,
piecewise-continuous Hamiltonian with $H(t+T)=H(t)$, and choose $t=0$ as a fixed phase of the drive. The time-ordered propagator and Floquet operator are
\begin{equation}
\label{eq1}
 U(t,t_0)=\mathcal T\exp\!\left[-\frac{i}{\hbar}
 \int_{t_0}^{t}ds\,H(s)\right],
 \quad U_F=U(T,0).
\end{equation}
Consider a normalized pure state evolving unitarily,
$\ket{\psi(t)}=U(t,0)\ket{\psi_0}$. Define its Fubini–Study angle from the initial ray by
\begin{equation}
 \mathcal L(t)=\arccos\!\left|\braket{\psi_0|\psi(t)}\right|,
 \qquad 0\leq\mathcal L(t)\leq\frac{\pi}{2}.
\end{equation}
Thus, for the return fidelity $F(t)=|\braket{\psi_0|\psi(t)}|^2$,
$\mathcal L(t)=\arccos\sqrt{F(t)}$. The projective-speed derivation in Appendix~\ref{app:dtc_MT_derivation} gives
\begin{equation}
\mathcal L(\tau)
 \leq \frac{1}{\hbar}\int_0^\tau dt\,\Delta E(t)
 \label{eq:geometric_MT_general}
\end{equation}
Here $\Delta E(t)$ is the instantaneous energy dispersion. For a piecewise-continuous drive, the instantaneous projective-speed relation holds almost everywhere; the finite set of switching times has zero measure in the integral. It does not require $[H(t),H(t')]=0$. We define the path-averaged energy dispersion by
\begin{equation}
\label{eq4}
 \overline{\Delta E}(\tau)
 =\frac{1}{\tau}\int_0^\tau dt\,\Delta E(t).
\end{equation}
For the trajectory up to time $\tau$, $\tau_{\rm MT}$ is the physical lower-bound functional. Equality is possible only when the projective trajectory follows a length-minimizing geodesic between its endpoints. For the periodic drive, evaluating Eqs.~\eqref{eq:geometric_MT_general} and \eqref{eq4} at the stroboscopic endpoint $t_n=nT$ gives the MT inequality
\begin{equation}
{
nT\geq \tau_{\rm MT}^{(n)},
\qquad
\tau_{\rm MT}^{(n)}
=
\frac{\hbar\mathcal L_n}
{\overline{\Delta E}(nT)}.
}
\label{eq:stroboscopic_MT_general}
\end{equation}
The denominator is the path-averaged energy dispersion of the physical instantaneous Hamiltonian over the full continuous-time evolution, including micromotion within each period. It is not computed from the quasienergy spectrum or from a branch-dependent Floquet Hamiltonian.

\subsection{Binary Floquet drive}
\label{subsec:binary_MT}
For a binary protocol, we take
\begin{equation}
H(t)=
\begin{cases}
H_1,&jT\leq t<jT+t_1,\\[2pt]
H_2,&jT+t_1\leq t<(j+1)T,
\end{cases}
\qquad T=t_1+t_2.
\end{equation}
The segment propagators and Floquet unitary are
\begin{equation}
U_1=e^{-iH_1t_1/\hbar},\quad U_2=e^{-iH_2t_2/\hbar},\quad U_F=U_2U_1.
\end{equation}
We introduce the boundary and intermediate states
\begin{equation}
\ket{\psi_{j,A}}=U_F^j\ket{\psi_0},
\qquad
\ket{\psi_{j,B}}=U_1\ket{\psi_{j,A}}
\end{equation}
For a state $\ket{\phi}$, write
$\Delta_\phi(O)=
\sqrt{\bra{\phi}O^2\ket{\phi}-\bra{\phi}O\ket{\phi}^2}$.
Because the first two moments of $H_\alpha$ are conserved within segment $\alpha$, the time-integrated energy dispersion is
\begin{equation}
\int_0^{nT}dt\,\Delta E(t)=
\sum_{j=0}^{n-1}
\left[
t_1\Delta_{\psi_{j,A}}(H_1)
+
t_2\Delta_{\psi_{j,B}}(H_2)
\right].
\end{equation}
To evaluate both contributions on the boundary state, define $H_2^{\langle B\rangle} =U_1^\dagger H_2U_1$. Unitarity gives
$\Delta_{\psi_{j,B}}(H_2)= \Delta_{\psi_{j,A}}(H_2^{\langle B\rangle})$, and hence
\begin{equation}
{
\tau_{\rm MT}^{(n)}
=
\frac{\hbar\mathcal L_n\,nT}{
\displaystyle
\sum_{j=0}^{n-1}
\left[
t_1\Delta_{\psi_{j,A}}(H_1)
+
t_2\Delta_{\psi_{j,A}}
\!\left(H_2^{\langle B\rangle}\right)
\right]}.
}
\label{eq:binary_MT_main}
\end{equation}
Appendix~\ref{app:dtc_MT_derivation} derives the conservation of the segment moments and the dressed-operator identity. Section~\ref{sec:dtc_MT} evaluates the dressed operator and Eq.~\eqref{eq:binary_MT_main} for the disordered Ising drive.

\section{Disordered Floquet Ising model}
\label{sec:dtc_MT}
We apply Eq.~\eqref{eq:binary_MT_main} to a disordered Ising chain with open boundary conditions. The two segment Hamiltonians are
\begin{align}
& H_1 =
\sum_{i=1}^{L-1}
J_{z,i}\sigma_i^z\sigma_{i+1}^z
+
\sum_{i=1}^{L}h_i\sigma_i^z,
\label{eq:DTC_H1_main}\\
& H_2(\epsilon) =
\Omega(\epsilon)
\sum_{i=1}^{L}\sigma_i^x.
\label{eq:DTC_H2_main}
\end{align}
Here $\sigma_i^\mu$ are Pauli operators with eigenvalues $\pm1$. Within each realization, the couplings $J_{z,i}$ and longitudinal fields $h_i$ are quenched and sampled as
\begin{equation}
\begin{aligned}
J_{z,i}&=J_z(1+0.2\eta_i),
&\eta_i&\sim\mathcal U[-1,1],\\
h_i&=4\pi\frac{\hbar}{T}\zeta_i,
&\zeta_i&\sim\mathcal U[0,1].
\end{aligned}
\end{equation}
We use $J_zT/\hbar$ as the dimensionless interaction coordinate. We take equal segment durations and the pulse convention
$t_1=t_2=\frac{T}{2}$ and $\Omega(\epsilon)T/(2\hbar)=\frac{\pi}{2}(1+\epsilon)$. Thus $\epsilon=0$ gives an ideal global $\pi$ rotation about the $x$ axis, whereas $\epsilon>0$ denotes an overrotation. Here $\epsilon$ is not a drive-frequency detuning. The segment and Floquet unitaries are
\begin{equation}
\begin{aligned}
U_1&=e^{-iH_1T/(2\hbar)},
&U_2(\epsilon)&=e^{-iH_2(\epsilon)T/(2\hbar)},\\
U_F(\epsilon)&=U_2(\epsilon)U_1.
\end{aligned}
\end{equation}
At the ideal pulse, $U_2(0)
=(-i)^L\prod_{i=1}^{L}\sigma_i^x$, where $(-i)^L$ is a global phase. The initial state is the N\'eel product state
\begin{equation}
\sigma_i^z\ket{\psi_0}=m_i^{(0)}\ket{\psi_0},
\qquad
m_i^{(0)}=(-1)^{i-1}.
\end{equation}
The locked period-two spin response defined in
Eq.~\eqref{eq:app_locked_response_window} is used only as a reference diagnostic and does not enter the MT relation. For the model in Eqs.~\eqref{eq:DTC_H1_main} and
\eqref{eq:DTC_H2_main}, define $K_i=h_i+J_{z,i-1}\sigma_{i-1}^z+J_{z,i}\sigma_{i+1}^z$,
where terms involving absent neighbors are omitted at $i=1$ and $i=L$. Each $K_i$ is a diagonal operator on the neighboring spins and commutes with $\sigma_i^x$ and $\sigma_i^y$. The first-segment evolution dresses the
transverse generator according to
\begin{equation}
H_2^{\langle B\rangle}
=\Omega(\epsilon)\sum_{i=1}^{L}\left[\sigma_i^x\cos\!\left(\frac{K_iT}{\hbar}\right)-\sigma_i^y\sin\!\left(\frac{K_iT}{\hbar}\right)\right].
\label{eq:dressed_H2_DTC_main}
\end{equation}
Appendix~\ref{app:dtc_MT_derivation} derives this conjugation identity. For equal segment durations, the path-averaged energy dispersion is:
\begin{equation}
\overline{\Delta E}(nT)
=\frac{1}{2n}
\sum_{j=0}^{n-1}\left[\Delta_{\psi_{j,A}}(H_1)+
\Delta_{\psi_{j,A}}
\!\left(H_2^{\langle B\rangle}\right)
\right],
\label{eq:DTC_average_dispersion_main}
\end{equation}
where $\ket{\psi_{j,A}}=U_F^j\ket{\psi_0}$. Denote the sum in Eq.~\eqref{eq:DTC_average_dispersion_main} by
\begin{align}
\mathcal D_n
&=\sum_{j=0}^{n-1}
\left[\Delta_{\psi_{j,A}}(H_1)+
\Delta_{\psi_{j,A}}\!\left(H_2^{\langle B\rangle}\right)\right].
\end{align}
Substitution into the general MT relation gives
\begin{equation}
{\tau_{\rm MT}^{(n)}
=\frac{2n\hbar\mathcal L_n}{\mathcal D_n},
\qquad nT\geq \tau_{\rm MT}^{(n)}.
}
\label{eq:DTC_MT_main}
\end{equation}

$\mathcal L_n$ is the endpoint Fubini–Study angle. The $H_1$ term is the energy dispersion of the Ising interaction and longitudinal fields at the start of each period. The dressed $H_2$ term is the transverse-pulse dispersion after the preceding $H_1$ segment. Because $H_2^{\langle B\rangle}$ contains the operator-valued $K_i$, it depends on both the disordered Ising couplings and the longitudinal fields. The parameter $J_z$ enters explicitly through $H_1$ and $K_i$, whereas $\epsilon$ enters through $\Omega(\epsilon)$. Both parameters also modify the orbit $U_F^j\ket{\psi_0}$. The variance of $H_2^{\langle B\rangle}$ is the variance of the complete many-body sum in Eq.~\eqref{eq:dressed_H2_DTC_main}. It includes intersite covariance terms and cannot, in general, be reduced to a sum of single-site variances. The physical MT functional is formed separately for each disorder realization before disorder averaging
\begin{equation}
\left[\tau_{\rm MT}^{(n)}\right]_{\rm dis}
=\left[
\frac{\hbar\mathcal L_n}
{\overline{\Delta E}(nT)}
\right]_{\rm dis},
\end{equation}
which is not generally equal to the ratio of the disorder-averaged numerator and denominator.

\section{Size normalization of the Floquet Mandelstam–Tamm diagnostic}
\label{sec:rescaled_mt}

For each disorder realization \(s\) and stroboscopic endpoint
\(t_n=nT\), \(n\geq1\), the time-dependent Mandelstam–Tamm (MT)
relation gives
\begin{equation}
 nT\geq \tau_{{\rm MT},n,L}^{(s)},
 \qquad
 \tau_{{\rm MT},n,L}^{(s)}
 =
 \frac{\hbar\mathcal L_{n,L}^{(s)}}
 {\overline{\Delta E}_{n,L}^{(s)}},
 \label{eq:rescaled_mt_sample_bound}
\end{equation}
where
\begin{align}
 \mathcal L_{n,L}^{(s)}
 &=
 \arccos\!\left|
 \langle\psi_0|\psi_n^{(s)}\rangle
 \right|,
 \label{eq:rescaled_mt_angle}\\
 \overline{\Delta E}_{n,L}^{(s)}
 &=
 \frac{1}{nT}\int_0^{nT}dt\,
 \Delta E_L^{(s)}(t),
 \label{eq:rescaled_mt_path_dispersion}\\
 \Delta E_L^{(s)}(t)
 &=
 \sqrt{
 \langle H_L^2(t)\rangle_{s,t}
 -
 \langle H_L(t)\rangle_{s,t}^{\,2}
 }.
 \nonumber
\end{align}
The numerator depends only on the selected endpoint, whereas the denominator
averages the instantaneous energy dispersion over the complete micromotion
from \(0\) to \(nT\).  Equation
\eqref{eq:rescaled_mt_sample_bound} is evaluated separately for every
disorder realization before averaging.  All trajectories retained in the
numerical analysis have
\(\overline{\Delta E}_{n,L}^{(s)}>0\).  If this quantity were zero, the
geometric MT relation would require
\(\mathcal L_{n,L}^{(s)}=0\), and the evolution would produce no nontrivial
MT ratio.  We retain \(T\) and \(\hbar\) in the analytic expressions and set
\(T=\hbar=1\) only in the numerical calculations.

\subsection{Scaling definition and covariance criterion}

We define the normalized path-averaged energy dispersion
\begin{equation}
 \sigma_{n,L}^{(s)}
 :=
 \frac{\overline{\Delta E}_{n,L}^{(s)}}{\sqrt L}
 \label{eq:rescaled_mt_sigma}
\end{equation}
and the corresponding size-normalized MT quantity
\begin{equation}
 \widetilde\tau_{{\rm MT},n,L}^{(s)}
 :=
 \sqrt L\,\tau_{{\rm MT},n,L}^{(s)}
 =
 \frac{\hbar\mathcal L_{n,L}^{(s)}}
 {\sigma_{n,L}^{(s)}}.
 \label{eq:rescaled_mt_definition}
\end{equation}
The dimensionless diagnostic used below is the disorder average of the
realization-wise ratio,
\begin{equation}
 \mathcal T_{n,L}(J_z,\epsilon)
 :=
 \left[
 \frac{\widetilde\tau_{{\rm MT},n,L}^{(s)}}{T}
 \right]_{\rm dis}
 =
 \left[
 \frac{\hbar\mathcal L_{n,L}^{(s)}}
 {T\sigma_{n,L}^{(s)}}
 \right]_{\rm dis}.
 \label{eq:calT_definition}
\end{equation}
The ratio is formed before disorder averaging because, in general,
\[
 \left[\frac{\mathcal L}{\sigma}\right]_{\rm dis}
 \neq
 \frac{[\mathcal L]_{\rm dis}}{[\sigma]_{\rm dis}}.
\]
The rescaling does not define an additional lower bound on the physical
elapsed time; multiplication of the samplewise MT inequality by \(\sqrt L\)
gives only
\begin{equation}
 \sqrt L\,nT
 \geq
 \widetilde\tau_{{\rm MT},n,L}^{(s)}.
 \label{eq:rescaled_mt_inequality}
\end{equation}

The \(\sqrt L\) normalization is motivated by the fluctuation scaling of a
local many-body Hamiltonian.  On segment \(\alpha\), write
\[
 H_{\alpha,L}
 =
 \sum_{x=1}^{M_L}\mathfrak h_{\alpha,x},
 \qquad
 M_L\leq \kappa L+O(1),
\]
where the local terms \(\mathfrak h_{\alpha,x}\) are uniformly bounded.  Defining
\begin{align}
 \delta\mathfrak h_{\alpha,x}^{(s)}(t)
 &=
 \mathfrak h_{\alpha,x}
 -
 \langle\mathfrak h_{\alpha,x}\rangle_{s,t},
 \nonumber\\
 C_{\alpha,xy}^{(s)}(t)
 &=
 \frac{1}{2}
 \left\langle
 \left\{
 \delta\mathfrak h_{\alpha,x}^{(s)}(t),
 \delta\mathfrak h_{\alpha,y}^{(s)}(t)
 \right\}
 \right\rangle_{s,t},
 \label{eq:local_covariance}
\end{align}
the instantaneous variance is
\begin{equation}
 \Delta^2H_{\alpha,L}^{(s)}(t)
 =
 \sum_{x,y}C_{\alpha,xy}^{(s)}(t).
 \label{eq:variance_covariance_sum}
\end{equation}
If the connected covariances are uniformly summable over the sizes,
realizations, drive segments, and times in the observation window,
\begin{equation}
 \sup_{L,s,\alpha,t,x}
 \sum_y
 \left|C_{\alpha,xy}^{(s)}(t)\right|
 \leq C_*<\infty,
 \label{eq:covariance_summability}
\end{equation}
then
\[
 \Delta^2H_{\alpha,L}^{(s)}(t)
 \leq M_LC_*
 =O(L),
\]
and consequently
\(\overline{\Delta E}_{n,L}^{(s)}=O(\sqrt L)\).
This is an upper scaling bound.  Normal fluctuation scaling,
\(\overline{\Delta E}_{n,L}=\Theta(\sqrt L)\), additionally requires the
normalized path-averaged energy dispersion to remain finite and nonzero.

For any sequence of states for which \(\sigma_{n,L}\) remains finite and
bounded away from zero and \(\mathcal L_{n,L}\) remains bounded away from zero,
\begin{equation}
 \tau_{{\rm MT},n,L}
 =
 \Theta(L^{-1/2}),
 \qquad
 \widetilde\tau_{{\rm MT},n,L}
 =
 \Theta(1).
 \label{eq:rescaled_mt_asymptotic_conditions}
\end{equation}
Because the disorder configurations are generated independently at each
\(L\), the large-\(L\) tests below concern disorder-averaged quantities based
on independent ensembles at each size.  The covariance derivation and its conditions are
given in Appendix~\ref{app:mt_covariance}.

\subsection{Finite-size validation and fixed-endpoint large-L behavior}
\label{subsec:mt_normalization_validation}

The endpoint test is performed at \(n=100\) on the native
\(40\times40\) \(L=16\) grid, with the \(L=8,10,12\) records bilinearly
aligned from their native \(39\times39\) axes to the \(L=16\) axes, using
\begin{equation}
 \mathcal S=\{8,10,12,16\}.
 \label{eq:four_size_set}
\end{equation}
Each native or aligned parameter cell contains \(N_{\rm dis}=50\)
realization-resolved endpoint angles and path-averaged energy uncertainties.
At each parameter cell, the effective exponent
\(\gamma_{\mathcal S}\) is obtained from a linear fit of
\(\ln[\overline{\Delta E}_{100,L}]_{\rm dis}\) against \(\ln L\)
\begin{equation}
 [\overline{\Delta E}_{100,L}]_{\rm dis}
 \propto L^{\gamma_{\mathcal S}}.
 \label{eq:gamma_effective}
\end{equation}
Because it is extracted from four finite sizes,
\(\gamma_{\mathcal S}\) is used only to test consistency with the expected
exponent \(1/2\).

For a positive size-dependent quantity \(X_L\), we define its relative
range over the four sizes by
\begin{equation}
 \delta_{\mathcal S}X
 =
 \frac{
 \max_{L\in\mathcal S}X_L
 -
 \min_{L\in\mathcal S}X_L
 }{
 \operatorname{mean}_{L\in\mathcal S}X_L
 }.
 \label{eq:relative_size_range}
\end{equation}
The two largest sizes are compared using the symmetric relative difference
\begin{equation}
 d_{12,16}(X)
 =
 \frac{2|X_{16}-X_{12}|}
 {|X_{16}|+|X_{12}|}.
 \label{eq:largest_size_relative_difference}
\end{equation}

At fixed endpoint \(n\) and parameters \((J_z,\epsilon)\), define, whenever
the limit exists,
\begin{equation}
 \mathcal C_{n,\infty}(J_z,\epsilon)
 :=
 \lim_{L\rightarrow\infty}
 \mathcal T_{n,L}(J_z,\epsilon).
 \label{eq:thermodynamic_MT_coefficient}
\end{equation}
The exact relation
\[
 \left[
 \frac{\tau_{{\rm MT},n,L}^{(s)}}{T}
 \right]_{\rm dis}
 =
 \frac{\mathcal T_{n,L}}{\sqrt L}
\]
then implies
\begin{equation}
 \left[
 \frac{\tau_{{\rm MT},n,L}^{(s)}}{T}
 \right]_{\rm dis}
 =
 \frac{\mathcal C_{n,\infty}}{\sqrt L}
 +o(L^{-1/2}).
 \label{eq:thermodynamic_MT_scaling}
\end{equation}
This definition takes \(L\rightarrow\infty\) at fixed endpoint \(n\); no
long-time limit is implied.

To separate the bounded endpoint angle from the normalized dispersion, we
introduce the saturated-angle upper bound on \(\mathcal T_{n,L}\),
\begin{equation}
 C^{\rm sat}_{n,L}
 :=
 \left[
 \frac{\pi\hbar}
 {2T\sigma_{n,L}^{(s)}}
 \right]_{\rm dis}.
 \label{eq:Csat_definition}
\end{equation}
Since \(0\leq\mathcal L_{n,L}^{(s)}\leq\pi/2\),
\begin{equation}
 C^{\rm sat}_{n,L}-\mathcal T_{n,L}
 =
 \left[
 \frac{
 \hbar[\pi/2-\mathcal L_{n,L}^{(s)}]
 }{
 T\sigma_{n,L}^{(s)}
 }
 \right]_{\rm dis}
 \geq0.
 \label{eq:Csat_gap_identity}
\end{equation}
If this weighted angular-deficit term vanishes and
\(C^{\rm sat}_{n,L}\) converges to a finite value, then
\begin{equation}
 \mathcal C_{n,\infty}
 =
 \lim_{L\rightarrow\infty}
 C^{\rm sat}_{n,L}.
 \label{eq:thermodynamic_MT_saturated}
\end{equation}

We select candidate angle-saturated cells using the operational criterion
\begin{equation}
 [\mathcal L_{100,16}]_{\rm dis}
 >
 0.95\frac{\pi}{2}.
 \label{eq:operational_angle_saturation}
\end{equation}
The weighted correction in Eq.~\eqref{eq:Csat_gap_identity} is then
evaluated separately within this subset. The threshold is neither a phase criterion nor an assumption about the thermodynamic limit.  Bootstrap intervals for the angular correction and \(C^{\rm sat}_{100,L}\) are evaluated while holding fixed the subset selected from the original sample; the stability of the classification under resampling is reported in
Appendix~\ref{app:mt_scaling_bootstrap}.

\begin{figure*}[t]
 \centering
 \includegraphics[width=\textwidth]
 {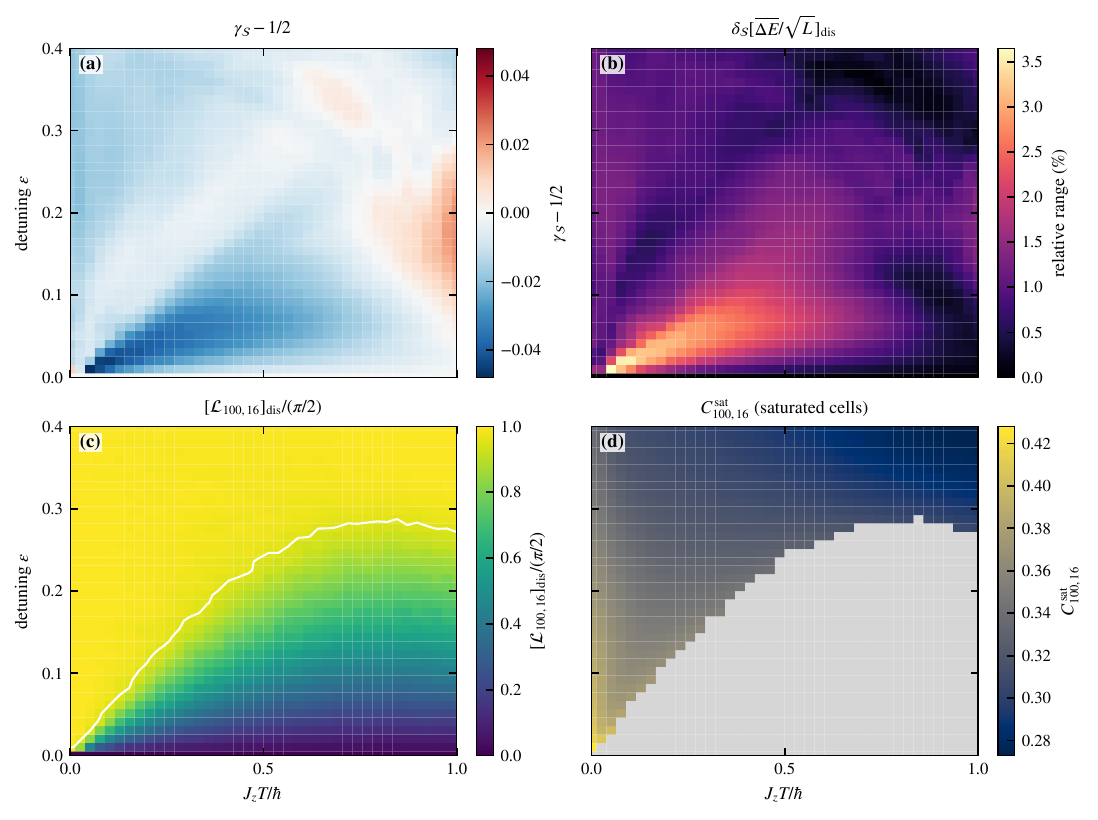}
 \caption{Four-size validation of the MT normalization at \(n=100\) on
 the native \(40\times40\) \(L=16\) parameter grid.  Each native or aligned
 cell contains
 \(N_{\rm dis}=50\) disorder realizations for
 \(L\in\mathcal S=\{8,10,12,16\}\).
 For the four-size quantities in panels (a) and (b), the \(L=8,10,12\)
 records are bilinearly aligned from their native \(39\times39\) axes to
 the \(L=16\) axes; panels (c) and (d) use native \(L=16\) values.  The
 aligned values are bilinear interpolants of the native records.
 (a) Signed deviation \(\gamma_{\mathcal S}-1/2\) of the effective dispersion exponent from normal-fluctuation scaling.  (b) Relative size range
 \(\delta_{\mathcal S}
 [\overline{\Delta E}_{100,L}/\sqrt L]_{\rm dis}\), expressed as a
 percentage.
 (c) Disorder-averaged endpoint angle at \(L=16\), normalized by
 \(\pi/2\); the white contour marks
 \([\mathcal L_{100,16}]_{\rm dis}=0.95(\pi/2)\).
 (d) Largest-size finite-\(L\) proxy
 \(C^{\rm sat}_{100,16}\) in the candidate angle-saturated subset;
 gray cells do not satisfy the selection criterion.
 No smoothing is applied; the only cross-grid operation is the stated
 bilinear alignment.  Statistical and correction-form controls are reported in
 Appendix~\ref{app:mt_scaling_bootstrap}.}
 \label{fig:mt_scaling_validation_revised}
\end{figure*}

The median of the cellwise effective-exponent estimates is \(0.492\), and
their 5th–95th percentile range is
\(0.472\)–\(0.506\)
[Fig.~\ref{fig:mt_scaling_validation_revised}(a)].  Bootstrap resampling
performed independently at each size gives a median of \(0.493\) and a 95\%
percentile interval \(0.447\)–\(0.537\) for the median exponent over the grid.  The
data are consistent with the normal-fluctuation value \(1/2\), although the
accessible sizes do not determine the exponent with high precision.

{The point estimate of the grid median of
\(\delta_{\mathcal S}
[\overline{\Delta E}_{100,L}/\sqrt L]_{\rm dis}\)
is approximately \(1.0\%\), with 95\% of the parameter cells below \(2.2\%\)
[Fig.~\ref{fig:mt_scaling_validation_revised}(b)].}  Under bootstrap
resampling performed independently at each size, the distribution of this nonnegative max–min
statistic has median \(2.4\%\) and a central 95\% interval
\(1.1\%\)–\(5.2\%\).  The increase relative to the raw point estimate
reflects the positive contribution of sampling fluctuations to a max–min
range.  These results support \(\sqrt L\) as the leading normalization of
the path-averaged energy dispersion over the accessible sizes and at the
present few-percent resolution; they do not imply a sub-percent size
collapse.

The angle criterion in Eq.~\eqref{eq:operational_angle_saturation} selects
795 of the 1560 cells with \(\epsilon>0\)
[Fig.~\ref{fig:mt_scaling_validation_revised}(c)].
{In this subset, the exact relative angular
correction
\(1-\mathcal T_{100,16}/C^{\rm sat}_{100,16}\)
has a median of \(1.0\%\) and a 95th percentile of \(3.8\%\) over the
parameter cells.}  Holding the selected subset fixed, the 95\% bootstrap percentile
interval for its grid median is \(0.92\%\)–\(1.08\%\).  The angle-only
criterion, defined without reference to \(\sigma_{100,16}^{(s)}\),
therefore identifies cells in which the weighted correction entering
Eq.~\eqref{eq:Csat_gap_identity} is numerically small.  Realization-level
influence and lower-tail checks reported in
Appendix~\ref{app:mt_scaling_bootstrap} confirm that the inverse-dispersion
average is not dominated by rare low-\(\sigma\) samples.

The raw point estimate of the median
\(d_{12,16}(C^{\rm sat}_{100,L})\) is \(0.2\%\).  Holding the selected
subset fixed, the bootstrap distribution of this median symmetric relative
difference has median \(0.9\%\) and a central 95\% interval
\(0.2\%\)–\(3.2\%\).  The corresponding median signed difference between
\(L=12\) and \(L=16\) is compatible
with zero, as detailed in Appendix~\ref{app:mt_scaling_bootstrap}.  Thus no
systematic median displacement between \(L=12\) and \(L=16\) is resolved,
while the statistical precision remains at the few-percent level.

The small weighted angular correction and the
absence of a resolved systematic median displacement between \(L=12\) and
\(L=16\) are consistent with a finite, nonzero fixed-endpoint limit in the
candidate angle-saturated subset.  We therefore use
\(C^{\rm sat}_{100,16}\), shown in
Fig.~\ref{fig:mt_scaling_validation_revised}(d), as a proxy for its amplitude
at the largest accessible size.  {Fits of the complete
diagnostic to \(C+aL^{-p}\), with \(p=1/2,1,2\), give a median intercept
spread of \(12.8\%\) and a 95th percentile of \(19.5\%\) within the
same subset.}  {This correction-form dependence is
the dominant finite-size sensitivity among the controls considered here and
prevents a controlled pointwise extrapolation over the parameter plane
\cite{Campostrini2014,Aramthottil2021}.}  The bootstrap construction and
correction-form analysis are detailed in
Appendix~\ref{app:mt_scaling_bootstrap}.  Time-dependent normalization tests
use \(L=8,10,12\), for which complete trajectories are available.

{Within the angle-saturated subset, the relative angular
correction \(1-\mathcal T_{100,16}/C^{\rm sat}_{100,16}\) remains at the
few-percent level. The variation of \(\mathcal T_{100,16}\) across this
subset is therefore governed predominantly by the normalized path-averaged
energy dispersion, as represented by \(C^{\rm sat}_{100,16}\) in
Fig.~\ref{fig:mt_scaling_validation_revised}(d).
Equation~\eqref{eq:DTC_average_dispersion_main} expresses this dispersion
through the standard deviations of \(H_1\) and the dressed pulse generator
\(H_2^{\langle B\rangle}\) evaluated along the stroboscopic orbit. At fixed
\(n\) and \(L\), \(\sigma_{n,L}^{(s)}\) is proportional to the
Fubini--Study length of the trajectory from \(0\) to \(nT\).}

The ideal-pulse line \(\epsilon=0\) provides an independent analytic check.
Let \(P=\prod_i\sigma_i^x\) and write \(H_1=H_J+H_h\).  At
\(\Omega T/(2\hbar)=\pi/2\),
\begin{equation}
 U_F^2
 =
 (-1)^L e^{-iH_JT/\hbar}.
 \label{eq:ideal_two_period_unitary}
\end{equation}
For the N\'eel initial state, every even endpoint returns to the initial
ray, whereas every odd endpoint is the orthogonal globally spin-reversed
state.  The normalized path-averaged energy dispersion and rescaled MT sequence are
\begin{equation}
 \sigma_{n,L}
 =
 \frac{\pi\hbar}{2T},
 \qquad
 \frac{\widetilde\tau_{{\rm MT},n,L}}{T}
 =
 \begin{cases}
 1, & n\ \mathrm{odd},\\
 0, & n\ \mathrm{even}.
 \end{cases}
 \label{eq:ideal_mt_sequence}
\end{equation}
Appendix~\ref{app:mt_ideal_pulse} gives the segment-resolved derivation.
Because the same recurrence occurs at \(J_z=0\), this exact sequence is an
implementation and normalization check rather than a criterion for
interaction-stabilized DTC order.

The subsequent phase-plane analysis therefore uses
\(\mathcal T_{n,L}\) as a size-normalized dynamical diagnostic.  The
normalization factors out the observed leading \(\sqrt L\) dependence of
the path-averaged energy dispersion while retaining the endpoint geometry.
Its relation to the independently determined finite-size dynamical regimes
is examined without identifying it as an additional physical-time bound or
a thermodynamic order parameter.


\section{Finite-size and phase-resolved structure of the rescaled MT diagnostic}
\label{sec:mt_results}

Section~\ref{sec:rescaled_mt} shows that the path-averaged energy
dispersion is consistent with leading $\sqrt{L}$ scaling over the
accessible sizes, while the complete MT ratio retains endpoint-angle
dependence and sensitivity to finite-size corrections.  The phase-resolved
analysis is restricted to the accessible finite sizes, leaving pointwise
thermodynamic extrapolation unresolved.  The results use the raw
$39\times39$ dynamical grids for $L=8,10,12$ in
$(J_zT/\hbar,\epsilon)$, with $N_{\rm dis}=50$ disorder realizations and an
observation window of $N=100$ periods.  The $L=16$ record contains only the
$n=100$ endpoint quantities and therefore does not enter the time-resolved
or response analyses.  Comparisons with quasienergy statistics use the
nested $20\times20$ spectral grid described in
Appendix~\ref{app:numerical_data_scope}.

\subsection{Endpoint parity and locked MT component}
\label{subsec:mt_endpoint_parity}

\begin{figure*}[t]
 \centering
 \includegraphics[width=\textwidth]
 {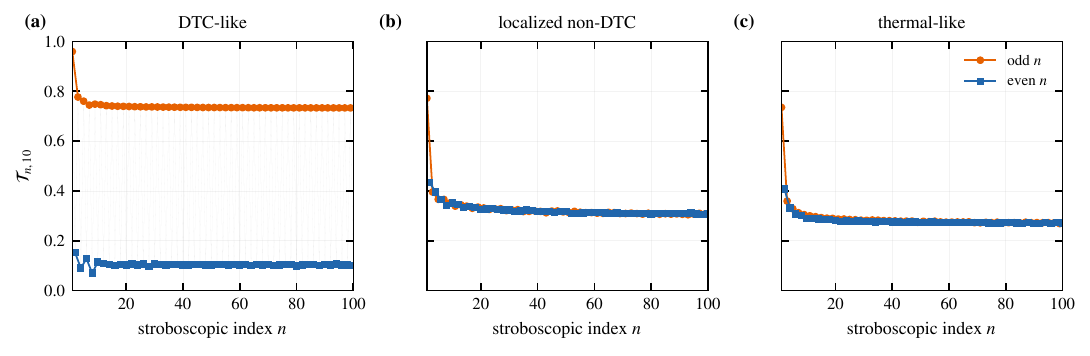}
 \caption{Endpoint dependence of the size-normalized MT diagnostic
 $\mathcal T_{n,10}$ at representative DTC-like, localized non-DTC, and
 thermal-like cells.  Their coordinates
 $(J_zT/\hbar,\epsilon)$ are $(0.789,0.042)$, $(0.211,0.295)$, and
 $(0.789,0.358)$, respectively.  Orange circles and blue squares denote
 odd and even endpoints.  Symbols show disorder means, lines join
 consecutive endpoints, and shaded bands show standard errors for
 $N_{\rm dis}=50$.  All panels use identical axes.  The sector labels refer
 to the finite-size, finite-time classification specified in
 Appendix~\ref{app:numerical_data_scope}.}
 \label{fig:mt_endpoint_parity_revised}
\end{figure*}

Because the MT numerator is a return distance from the initial ray,
$\mathcal T_{n,L}$ can depend strongly on the parity of the stroboscopic
endpoint.  Figure~\ref{fig:mt_endpoint_parity_revised} illustrates this
dependence at $L=10$ for representative DTC-like, localized non-DTC, and
thermal-like cells.  The points were selected from the nested
grid within the response- and spectrum-based operational sectors defined
independently of the MT diagnostic.  The selection procedure is specified
in Appendix~\ref{app:mt_result_controls}.

{At the representative DTC-like cell, the odd and
even branches remain separated throughout the available 100-period record.}
{Their standard-error bands remain disjoint at the
last sampled odd and even endpoints.}
{The localized non-DTC and thermal-like comparison
cells exhibit nearly coincident late-time branches
[Fig.~\ref{fig:mt_endpoint_parity_revised}].}

The ideal-pulse line supplies an exact limiting check.  For $\epsilon=0$
and the N\'eel initial state, Appendix~\ref{app:mt_ideal_pulse} gives
\begin{equation}
 \mathcal T_{n,L}
 =
 \begin{cases}
 1, & n\ \text{odd},\\
 0, & n\ \text{even},
 \end{cases}
 \label{eq:results_ideal_parity}
\end{equation}
independently of $J_z$.  This exact sequence results from endpoint
recurrence and also occurs in the noninteracting limit.  It is therefore
not, by itself, evidence for interaction-stabilized DTC order.

To characterize parity locking without selecting a single endpoint, we
define
\begin{equation}
 m_{\pi,L}^{\rm MT}
 =
 \left[
 \left|
 \frac{1}{N}
 \sum_{n=1}^{N}
 (-1)^n
 \frac{\widetilde{\tau}_{{\rm MT},n,L}^{(s)}}{T}
 \right|
 \right]_{\rm dis}.
 \label{eq:results_locked_mt}
\end{equation}
The absolute value is taken realization by realization before disorder
averaging, as in the definition of the locked spin response.  Thus
$m_{\pi,L}^{\rm MT}$ is the amplitude of the period-two component of the
sequence of endpoint MT quantities.  It is a dimensionless dynamical
diagnostic, not a lower bound on an elapsed time.

The distinction is quantitatively important.  At $n=100$,
$\tau_{\rm MT}/(100T)$ remains below $2.1\times10^{-3}$ throughout the
available data, so the underlying MT inequality is far from saturation.
The parity structure of $m_{\pi,L}^{\rm MT}$ therefore characterizes the
endpoint sequence rather than the tightness of the speed limit.  The
separate angle, normalized-dispersion, and MT-ratio histories at the
representative DTC-like cell are shown in
Fig.~\ref{fig:app_mt_time_components}.

{The time-resolved decomposition in
Fig.~\ref{fig:app_mt_time_components} shows strong odd--even modulation of
the disorder-averaged endpoint angle. The disorder-averaged normalized path
dispersion shows little discernible odd--even modulation after the initial
transient. Endpoint-angle alternation therefore accounts for most of the
period-two MT component at this representative DTC-like point. Since return
fidelity determines the endpoint angle, it captures the dominant
contribution to this component.}

\subsection{Comparison with response and spectral crossovers}
\label{subsec:mt_phase_maps}

\begin{figure*}[t]
 \centering
 \includegraphics[width=\textwidth]
 {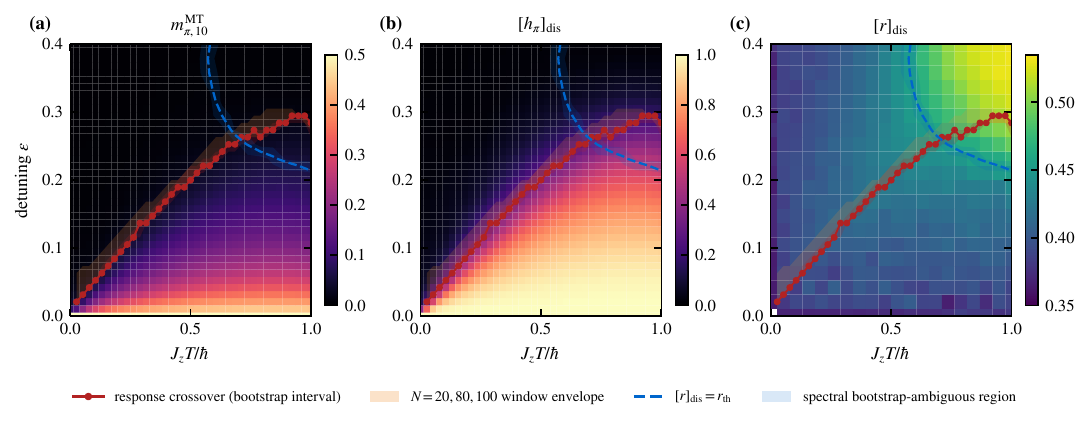}
 \caption{Finite-size dynamical and spectral maps at $L=10$, with
 $N=100$ and $N_{\rm dis}=50$.
 (a) Locked MT component $m_{\pi,10}^{\rm MT}$.
 (b) Disorder-averaged locked spin response $[h_\pi]_{\rm dis}$, where
 $h_\pi=h_{\pi,s,10}^{[1,100]}$ before disorder averaging.
 Panels (a) and (b) show the raw $39\times39$ dynamical grid.
 (c) Disorder-averaged adjacent-gap ratio $[r]_{\rm dis}$ on the
 available raw $20\times20$ spectral grid.
 Red markers show the response-variance crossover estimator for the
 $[1,100]$ window; the dark-red band is its 16th–84th percentile
 disorder-bootstrap interval from 2000 resamples.  The orange band is the
 envelope obtained from the available $N=20,80,100$ observation windows.
 The dashed blue curve marks $[r]_{\rm dis}=0.4565$, and the light-blue
 region contains cells satisfying
 $0.16\leq P_{\rm boot}([r]_{\rm dis}>0.4565)\leq0.84$.
 Heat-map cells are neither interpolated nor smoothed.  Lines joining grid
 centers are guides to the eye; all overlays are operational finite-size
 crossover estimators.}
 \label{fig:mt_phase_association_revised}
\end{figure*}

Figure~\ref{fig:mt_phase_association_revised} compares the locked MT
component with the independently defined spin-response diagnostic and with
quasienergy statistics obtained from a separate spectral calculation at
$L=10$.  The parity behavior in
Fig.~\ref{fig:mt_endpoint_parity_revised} motivates the projection used in
panel (a).  Panels (b) and (c) show, respectively, the locked spin response
and the adjacent-gap ratio on the same parameter plane.

The locked MT component is largest near the ideal-pulse line and generally
decreases with increasing pulse error
[Fig.~\ref{fig:mt_phase_association_revised}(a)].  Its phase-plane
variation closely tracks that of the locked spin response in
Fig.~\ref{fig:mt_phase_association_revised}(b).  On the displayed
finite-size grid, both maps exhibit enhanced period-two weight on the
low-pulse-error side of the response crossover, and the extent of this region
increases with $J_z$.  The bright row at $\epsilon=0$ in both maps reflects the
exact $J_z$-independent recurrence in Eq.~\eqref{eq:results_ideal_parity} and
therefore does not distinguish an interaction-stabilized DTC response from a
noninteracting perfect pulse.

The red curve identifies, at each $J_z$, the pulse error at which the disorder
variance of the locked spin response is maximal.  The dark-red band
quantifies disorder-sampling uncertainty, while the orange band records the
variation among the available observation windows.  These uncertainties
are kept separate because bootstrap sampling and changing the observation
time probe distinct sources of variation.  At fixed observation window, the
reported disorder-bootstrap interval quantifies disorder-sampling variation of
the response-crossover coordinate at the available grid resolution;
displacement among the sampled time windows quantifies its observation-window
dependence.
Quantitative grid-width and bootstrap controls are reported in
Appendix~\ref{app:mt_result_controls}.

The adjacent-gap-ratio map in
Fig.~\ref{fig:mt_phase_association_revised}(c) uses the $L=10$
quasienergy data.  {Its contour
$[r]_{\rm dis}=r_{\rm th}$ marks the $L=10$ operational crossover between
localized-like and thermal-like spectral statistics defined in
Appendix~\ref{app:mt_spectral_comparison}.}
The corresponding $L=8$ comparison in
Fig.~\ref{fig:app_mt_spectral_comparison} displays a finite displacement of
this contour between the two sizes.  The data therefore identify
finite-size DTC-like, localized non-DTC, and thermal-like sectors but do not
determine a thermodynamic three-phase diagram.

No abrupt or uniquely identifiable feature of $m_{\pi,10}^{\rm MT}$ follows
the complete spectral crossover.  Its relationship to quasienergy
statistics must consequently be distinguished from its close association
with the subharmonic response.

\subsection{Association tests}
\label{subsec:mt_association_tests}

\begin{figure*}[h]
 \centering
 \includegraphics[width=\textwidth]
 {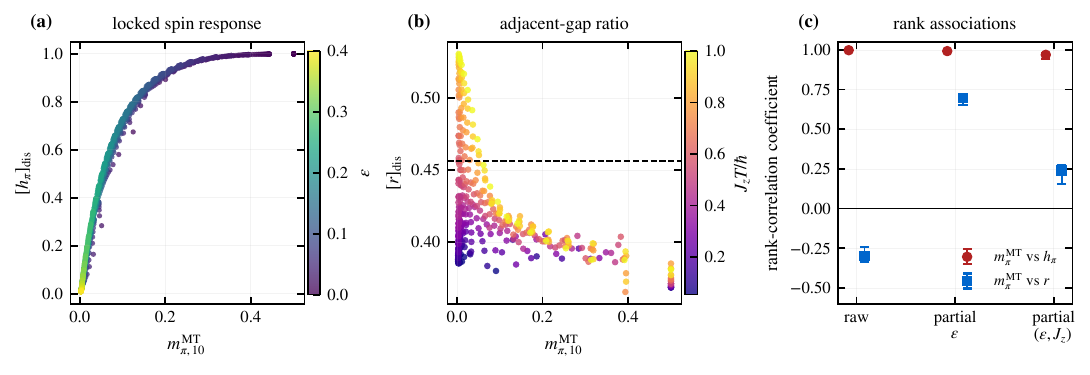}
 \caption{Association tests at $L=10$.
 (a) Locked MT component $m_{\pi,10}^{\rm MT}$ versus the
 disorder-averaged locked spin response $[h_\pi]_{\rm dis}$ for the 1482
 cells with $J_z>0$ on the $39\times39$ dynamical grid; color denotes
 pulse error $\epsilon$.
 (b) Locked MT component versus the disorder-averaged adjacent-gap
 ratio $[r]_{\rm dis}$ for the 380 cells with $J_z>0$ on the nested
 $20\times20$ spectral grid; color denotes $J_zT/\hbar$, and the dashed
 line marks $r_{\rm th}=0.4565$.
 (c) Raw and partial Spearman rank coefficients.  Partial coefficients are
 obtained after linear residualization of the ranked observables against
 the ranked coordinates indicated on the horizontal axis.  Error bars are
 95\% percentile intervals from 2000 realization-bootstrap resamples.
 Dynamical quantities are resampled jointly, whereas the independently
 generated dynamical and spectral ensembles are resampled separately.
 These coefficients describe associations over the sampled parameter grids
 and are not significance tests based on the number of grid cells.}
 \label{fig:mt_association_scatter}
\end{figure*}

The visual comparisons in
Fig.~\ref{fig:mt_phase_association_revised} are quantified in
Fig.~\ref{fig:mt_association_scatter}.  Spearman rank coefficients are
calculated over the 1482 interacting cells of the $39\times39$ dynamical
grid for the MT–response comparison and over the 380 interacting cells of
the nested $20\times20$ grid for the MT–spectral comparison.  The
noninteracting line $J_z=0$ is excluded from these association tests but
remains displayed in the phase-plane maps.

Partial-rank coefficients are constructed by rank transforming the two
observables and the indicated parameter coordinates, linearly
residualizing both observables against the ranked coordinates, and
correlating the residuals.  This procedure removes the coordinate
dependence captured by the linear rank-space regression; it does not remove
arbitrary nonlinear parameter dependence.  Bootstrap intervals are
obtained by resampling disorder realizations rather than by treating
neighboring parameter cells as independent measurements.

For $m_{\pi,10}^{\rm MT}$ and $[h_\pi]_{\rm dis}$, the raw Spearman
coefficient is $0.9981$, with a 95\% bootstrap interval
$0.9967$–$0.9984$.  It remains $0.9926$
($0.9853$–$0.9946$) after residualizing against the ranked pulse-error coordinate and
$0.9693$ ($0.9422$–$0.9788$) after residualizing jointly against the
ranked pulse-error and interaction coordinates.  The association therefore
survives removal of the monotonic coordinate dependence represented by
this partial-rank procedure.

The close correspondence shows that the locked component of the MT
endpoint sequence tracks the period-two structure detected by the spin
autocorrelation.  Both observables, however, are calculated from the same
realization-resolved trajectories and use the same $(-1)^n$ projection.
Their agreement is therefore not independent evidence for DTC order, and
$m_{\pi,L}^{\rm MT}$ is not an additional DTC order parameter.

The comparison with quasienergy statistics is qualitatively different.
The raw coefficient between $m_{\pi,10}^{\rm MT}$ and $[r]_{\rm dis}$ is
$-0.2993$, with a 95\% bootstrap interval
$-0.3338$ to $-0.2414$.  It becomes $0.6962$
($0.6515$–$0.7123$) after residualizing against the ranked pulse-error coordinate and
decreases to $0.2407$ ($0.1573$–$0.2761$) after residualizing jointly
against the ranked pulse-error and interaction coordinates.  The sign change
and strong dependence on the chosen coordinate residualization show that
the data do not support a single parameter-independent monotonic relation
between $m_{\pi,10}^{\rm MT}$ and $[r]_{\rm dis}$ over the sampled grid.

The rescaled MT sequence therefore provides a geometric and energetic
characterization of the period-two Floquet trajectory that complements the
spin response.  Quasienergy statistics retain their role in classifying
localized-like and thermal-like spectra.

\section{Conclusion}
\label{sec:conclusion}

We derived an exact segment-resolved form of the pure-state
Mandelstam–Tamm (MT) lower-bound functional for binary Floquet drives with
arbitrary positive segment durations, without assuming commutativity of the
segment Hamiltonians. For piecewise-constant protocols, the time integral of
the instantaneous energy dispersion is evaluated exactly segment by segment,
retaining intraperiod micromotion. In the disordered Floquet Ising chain,
conjugating the transverse-pulse Hamiltonian by the first-segment propagator
yields an explicit dressed operator that allows both segment contributions to
the MT denominator to be evaluated on the stroboscopic orbit.

For local segment Hamiltonians, absolute and uniform summability of connected
local-energy covariances implies an $O(\sqrt L)$ upper bound on the
path-averaged energy dispersion. The $n=100$ data for $L=8,10,12,16$ are
consistent with leading $\sqrt L$ scaling of this quantity over the accessible
sizes and support the corresponding rescaling.

{Within the angle-saturated subset at \(n=100\), the
normalized path-averaged energy dispersion accounts for most of the variation
of \(\mathcal T_{100,16}\) across this subset. Across the remaining cells, the
endpoint angle retains finite-size dependence, and the fitted intercept of the
rescaled MT diagnostic depends on the chosen correction exponent.}

In the finite-size region with a locked spin response, the period-two component
of the rescaled MT sequence closely follows the locked spin response. At the
representative DTC-like cell, the sequence remains separated into distinct odd
and even branches throughout the 100-period record. The endpoint angle accounts
for most of this separation, whereas temporal modulation of the normalized
path-averaged dispersion contributes only weakly. The tested rank-based controls
for pulse error and interaction strength preserve the close MT–response
association, whereas the MT–spectral association is strongly control
dependent. The MT–response correspondence links finite-time subharmonic
structure in the staggered-spin dynamics to the global return geometry of the
same trajectories.

Because the unrescaled MT bound remains far from saturation, the rescaled
quantity serves as a state-dependent diagnostic of the trajectory generated
from the specified N\'eel initial state. {The decomposition of
the size-rescaled MT diagnostic identifies the respective contributions of the
endpoint angle and normalized path-averaged energy dispersion. The endpoint
angle is determined by return fidelity, and at fixed elapsed time the
path-averaged dispersion determines the Fubini--Study length accumulated along
the trajectory. Autocorrelation and quasienergy statistics define the
dynamical and spectral regimes used in this comparison.}


\clearpage

\appendix
\section{Derivation of the Mandelstam–Tamm bound for the
disordered Floquet Ising model}
\label{app:dtc_MT_derivation}

This appendix derives the MT relation used in
Secs.~\ref{sec:floquet_mt} and \ref{sec:dtc_MT}.  We retain $\hbar$
explicitly and restrict the proof to normalized pure states undergoing
closed-system unitary evolution.  The finite spin chains considered here
have bounded Hamiltonians.  Piecewise-constant switching makes the state
trajectory piecewise differentiable, which is sufficient for the geometric
derivation.

\subsection{Instantaneous projective speed}

Let
\begin{equation}
i\hbar\ket{\dot\psi(t)}
=
H(t)\ket{\psi(t)},
\qquad
\braket{\psi(t)|\psi(t)}=1.
\label{eq:app_schrodinger}
\end{equation}
The Fubini–Study line element in the convention used in the main text is
\begin{equation}
d\ell_{\rm FS}^2
=
\braket{d\psi|d\psi}
-
\left|\braket{\psi|d\psi}\right|^2.
\label{eq:app_FS_line_element}
\end{equation}
Setting $\ket{d\psi}=\ket{\dot\psi}dt$ gives
\begin{equation}
\left(\frac{d\ell_{\rm FS}}{dt}\right)^2
=
\braket{\dot\psi|\dot\psi}
-
\left|\braket{\psi|\dot\psi}\right|^2.
\label{eq:app_FS_velocity_projected}
\end{equation}
The Schr\"odinger equation implies
\begin{equation}
\braket{\dot\psi|\dot\psi}
=
\frac{\bra{\psi}H(t)^2\ket{\psi}}{\hbar^2},
\qquad
\braket{\psi|\dot\psi}
=
-\frac{i}{\hbar}
\bra{\psi}H(t)\ket{\psi}.
\label{eq:app_velocity_moments}
\end{equation}
Therefore,
\begin{align}
\left(\frac{d\ell_{\rm FS}}{dt}\right)^2
&=
\frac{\bra{\psi(t)}H(t)^2\ket{\psi(t)}
-\bra{\psi(t)}H(t)\ket{\psi(t)}^2}{\hbar^2}
\nonumber\\
&=\frac{\Delta E(t)^2}{\hbar^2}.
\label{eq:app_FS_speed}
\end{align}
Taking the nonnegative square root gives
$d\ell_{\rm FS}/dt=\Delta E(t)/\hbar$.

For pure states, the geodesic distance between the initial and final rays is
\begin{equation}
\mathcal L(\tau)
=
\arccos\!\left|
\braket{\psi(0)|\psi(\tau)}
\right|.
\label{eq:app_geodesic_distance}
\end{equation}
The geodesic distance cannot exceed the length of the actual path
\begin{align}
\mathcal L(\tau)
&\leq
\int_0^\tau d\ell_{\rm FS}
\nonumber\\
&=
\frac{1}{\hbar}
\int_0^\tau dt\,\Delta E(t).
\label{eq:app_geodesic_path_inequality}
\end{align}
Defining
\begin{equation}
\overline{\Delta E}(\tau)
=
\frac{1}{\tau}
\int_0^\tau dt\,\Delta E(t),
\label{eq:app_average_dispersion}
\end{equation}
we obtain, for $\overline{\Delta E}(\tau)>0$,
\begin{equation}
\tau
\geq
\frac{\hbar\mathcal L(\tau)}
{\overline{\Delta E}(\tau)}.
\label{eq:app_MT_arbitrary}
\end{equation}

This proof does not assume $[H(t),H(t')]=0$. Non-commutativity changes the trajectory and its instantaneous dispersion, but not the projective path-length inequality. The switching points of a piecewise-constant drive form a set of zero measure and do not contribute separately to the integral. For a mixed state, the Bures metric involves the quantum Fisher information rather than the pure-state expression in Eq.~\eqref{eq:app_FS_line_element}. No mixed-state extension is used in the present work.

\subsection{Exact decomposition for a binary drive}

Consider
\begin{equation}
H(t)=
\begin{cases}
H_1,&jT\leq t<jT+t_1,\\[2pt]
H_2,&jT+t_1\leq t<(j+1)T,
\end{cases}
\qquad
T=t_1+t_2.
\label{eq:app_binary_drive}
\end{equation}
The segment propagators are
\begin{equation}
U_1=e^{-iH_1t_1/\hbar},
\qquad
U_2=e^{-iH_2t_2/\hbar},
\qquad
U_F=U_2U_1.
\label{eq:app_binary_unitaries}
\end{equation}
The boundary and intermediate states are
\begin{equation}
\ket{\psi_{j,A}}=U_F^j\ket{\psi_0},
\qquad
\ket{\psi_{j,B}}=U_1\ket{\psi_{j,A}}.
\label{eq:app_AB_states}
\end{equation}

During a segment generated by the time-independent Hamiltonian
$H_\alpha$, write $\langle O\rangle_t=
\bra{\psi(t)}O\ket{\psi(t)}$.  Its first two moments are conserved
\begin{equation}
\frac{d}{dt}
\langle H_\alpha^m\rangle_t
=\frac{i}{\hbar}\langle[H_\alpha,H_\alpha^m]\rangle_t=0,
\qquad m=1,2.
\label{eq:app_segment_moment_conservation}
\end{equation}
Both the mean and second moment are therefore constant, and so is
$\Delta(H_\alpha)$.  It follows that
\begin{align}
\int_{jT}^{jT+t_1}dt\,\Delta E(t)
&=
t_1\Delta_{\psi_{j,A}}(H_1),
\label{eq:app_first_segment_integral}\\
\int_{jT+t_1}^{(j+1)T}dt\,\Delta E(t)
&=
t_2\Delta_{\psi_{j,B}}(H_2).
\label{eq:app_second_segment_integral}
\end{align}
Summing over $n$ periods gives
\begin{align}
\mathcal I_{\Delta E}(nT)
&:=\int_0^{nT}dt\,\Delta E(t)\nonumber\\
&=\sum_{j=0}^{n-1}
\bigl[t_1\Delta_{\psi_{j,A}}(H_1)\nonumber\\
&\hspace{4.5em}+t_2\Delta_{\psi_{j,B}}(H_2)\bigr].
\label{eq:app_integrated_dispersion_binary}
\end{align}

Define
\begin{equation}
H_2^{\langle B\rangle}
=
U_1^\dagger H_2U_1.
\label{eq:app_dressed_H2}
\end{equation}
The first two moments in the intermediate state satisfy
\begin{align}
\bra{\psi_{j,B}}H_2\ket{\psi_{j,B}}
&=
\bra{\psi_{j,A}}
H_2^{\langle B\rangle}
\ket{\psi_{j,A}},
\label{eq:app_dressed_first_moment}\\
\bra{\psi_{j,B}}H_2^2\ket{\psi_{j,B}}
&=
\bra{\psi_{j,A}}
U_1^\dagger H_2^2U_1
\ket{\psi_{j,A}}
\nonumber\\
&=
\bra{\psi_{j,A}}
\left(H_2^{\langle B\rangle}\right)^2
\ket{\psi_{j,A}}.
\label{eq:app_dressed_second_moment}
\end{align}
The second equality uses the unitarity identity
\begin{equation}
\left(U_1^\dagger H_2U_1\right)^2
=
U_1^\dagger H_2^2U_1.
\end{equation}
Consequently,
\begin{equation}
\Delta_{\psi_{j,B}}(H_2)
=
\Delta_{\psi_{j,A}}
\left(H_2^{\langle B\rangle}\right),
\label{eq:app_dressed_dispersion}
\end{equation}
and
\begin{equation}
\mathcal I_{\Delta E}(nT)
=
\sum_{j=0}^{n-1}
\left[
t_1\Delta_{\psi_{j,A}}(H_1)
+
t_2\Delta_{\psi_{j,A}}
\left(H_2^{\langle B\rangle}\right)
\right].
\label{eq:app_integrated_dispersion_dressed}
\end{equation}

\subsection{Dressed transverse generator}

For the disordered Ising chain with open boundary conditions,
\begin{align}
H_1
&=
\sum_{i=1}^{L-1}
J_{z,i}\sigma_i^z\sigma_{i+1}^z
+
\sum_{i=1}^{L}h_i\sigma_i^z,
\label{eq:app_Ising_H1}\\
H_2
&=
\Omega(\epsilon)
\sum_{i=1}^{L}\sigma_i^x.
\label{eq:app_Ising_H2}
\end{align}
Introduce
\begin{equation}
K_i
=
h_i
+
J_{z,i-1}\sigma_{i-1}^z
+
J_{z,i}\sigma_{i+1}^z,
\qquad
J_{z,0}=J_{z,L}=0.
\label{eq:app_Ki}
\end{equation}
Using
\begin{equation}
[\sigma_i^z,\sigma_i^x]=2i\sigma_i^y,
\qquad
[\sigma_i^z,\sigma_i^y]=-2i\sigma_i^x,
\label{eq:app_Pauli_commutators}
\end{equation}
we find
\begin{equation}
[H_1,\sigma_i^x]
=
2iK_i\sigma_i^y,
\qquad
[H_1,\sigma_i^y]
=
-2iK_i\sigma_i^x.
\label{eq:app_closed_commutators}
\end{equation}
Only terms in $H_1$ containing $\sigma_i^z$ contribute.  Since $K_i$
contains only the identity and neighboring $\sigma^z$ operators,
\begin{equation}
[K_i,H_1]=[K_i,\sigma_i^x]=[K_i,\sigma_i^y]=0.
\label{eq:app_Ki_commutation}
\end{equation}
The nested commutator series therefore closes
\begin{align}
U_1^\dagger\sigma_i^xU_1
&=
e^{iH_1t_1/\hbar}
\sigma_i^x
e^{-iH_1t_1/\hbar}
\nonumber\\
&=
\sigma_i^x
\cos\!\left(\frac{2K_it_1}{\hbar}\right)
-
\sigma_i^y
\sin\!\left(\frac{2K_it_1}{\hbar}\right),
\label{eq:app_rotated_sigmax}\\
U_1^\dagger\sigma_i^yU_1
&=
\sigma_i^y
\cos\!\left(\frac{2K_it_1}{\hbar}\right)
+
\sigma_i^x
\sin\!\left(\frac{2K_it_1}{\hbar}\right).
\label{eq:app_rotated_sigmay}
\end{align}
The minus sign in Eq.~\eqref{eq:app_rotated_sigmax} follows directly from
\begin{equation}
\left.
\frac{d}{dt}
\left(
e^{iH_1t/\hbar}\sigma_i^x e^{-iH_1t/\hbar}
\right)
\right|_{t=0}
=
\frac{i}{\hbar}[H_1,\sigma_i^x]
=
-\frac{2K_i}{\hbar}\sigma_i^y.
\label{eq:app_rotation_sign_check}
\end{equation}

For $t_1=T/2$,
\begin{equation}
U_1^\dagger\sigma_i^xU_1
=
\sigma_i^x
\cos\!\left(\frac{K_iT}{\hbar}\right)
-
\sigma_i^y
\sin\!\left(\frac{K_iT}{\hbar}\right).
\label{eq:app_rotated_sigmax_equal}
\end{equation}
Hence
\begin{equation}
{
H_2^{\langle B\rangle}
=
\Omega(\epsilon)
\sum_{i=1}^{L}
\left[
\sigma_i^x
\cos\!\left(\frac{K_iT}{\hbar}\right)
-
\sigma_i^y
\sin\!\left(\frac{K_iT}{\hbar}\right)
\right]
}.
\label{eq:app_dressed_H2_closed}
\end{equation}
The trigonometric functions are operator functions of the neighboring
$\sigma^z$ operators.  They commute with the corresponding
$\sigma_i^x$ and $\sigma_i^y$, so
Eq.~\eqref{eq:app_dressed_H2_closed} is Hermitian.  The derivation retains
$K_i$ as an operator throughout.

\subsection{Model-specific MT expression}

For $t_1=t_2=T/2$,
Eq.~\eqref{eq:app_integrated_dispersion_dressed} becomes
\begin{equation}
\mathcal I_{\Delta E}(nT)
=
\frac{T}{2}
\sum_{j=0}^{n-1}
\left[
\Delta_{\psi_{j,A}}(H_1)
+
\Delta_{\psi_{j,A}}
\left(H_2^{\langle B\rangle}\right)
\right].
\label{eq:app_DTC_integrated_dispersion}
\end{equation}
Dividing by $nT$ gives
\begin{equation}
\overline{\Delta E}(nT)
=
\frac{1}{2n}
\sum_{j=0}^{n-1}
\left[
\Delta_{\psi_{j,A}}(H_1)
+
\Delta_{\psi_{j,A}}
\left(H_2^{\langle B\rangle}\right)
\right].
\label{eq:app_DTC_average_dispersion}
\end{equation}

The return fidelity and pure-state angle at the endpoint are
\begin{equation}
F_n
=
\left|
\bra{\psi_0}U_F^n\ket{\psi_0}
\right|^2,
\qquad
\mathcal L_n
=
\arccos\sqrt{F_n}.
\label{eq:app_endpoint_fidelity}
\end{equation}
Substitution of Eq.~\eqref{eq:app_DTC_average_dispersion} into
Eq.~\eqref{eq:app_MT_arbitrary} yields
\begin{equation}
{
\tau_{\rm MT}^{(n)}
=\frac{\hbar\mathcal L_n}{\overline{\Delta E}(nT)},
\qquad nT\geq \tau_{\rm MT}^{(n)}
}.
\label{eq:app_DTC_MT_final}
\end{equation}
The denominator has units of energy, so
Eq.~\eqref{eq:app_DTC_MT_final} has units of time.  The factor $1/(2n)$ in
Eq.~\eqref{eq:app_DTC_average_dispersion} originates from dividing the
dispersion accumulated over the equal half-periods by the
total time $nT$.

No quasienergy appears in Eq.~\eqref{eq:app_DTC_MT_final}.  A Floquet
Hamiltonian may be defined by
\begin{equation}
U_F=e^{-iH_FT/\hbar},
\end{equation}
but the logarithm is branch dependent, and $H_F$ does not, by itself, retain
the micromotion contribution to the projective path length.  Replacing
$\Delta E(t)$ by $\Delta H_F$ would therefore define a different
quantity and would not reproduce the derivation above in general.

If $\overline{\Delta E}(nT)=0$, then the nonnegative function
$\Delta E(t)$ vanishes almost everywhere.  Equation~\eqref{eq:app_FS_speed}
then implies that the state changes only by an overall phase and
$\mathcal L_n=0$.  In this case the ratio defining
$\tau_{\rm MT}^{(n)}$ is not used; the geometric inequality is satisfied
trivially.

\section{Relation between the staggered spin response and the two-time autocorrelation}
\label{app:autocorrelation}

For disorder realization $s$, define the staggered spin response by
\begin{equation}
 \begin{aligned}
 R_{s,L}(nT)
 &=\frac{1}{L}\sum_{i=1}^{L}m_i^{(0)}
 \bra{\psi_{n,s}}\sigma_i^z\ket{\psi_{n,s}},\\
 \ket{\psi_{n,s}}&=U_{F,s}^{n}\ket{\psi_0}.
 \end{aligned}
 \label{eq:app_staggered_response}
\end{equation}
For a product initial state in the $\sigma^z$ basis, this response is exactly
the site-averaged stroboscopic two-time autocorrelation used in Floquet-DTC
studies \cite{Yao2017,Zhang2017,Khemani2016,vonKeyserlingk2016,Zaletel2023}.
The identity holds separately for every disorder realization; we omit the
realization label in the remainder of this section.

Let
\begin{equation}
 \ket{\psi_n}=U_F^n\ket{\psi_0},
 \qquad
 \sigma_i^z(nT)
 =\left(U_F^\dagger\right)^n\sigma_i^zU_F^n.
 \label{eq:app_heisenberg_spin}
\end{equation}
The local two-time autocorrelation is
\begin{equation}
 R_i(nT)
 =\bra{\psi_0}\sigma_i^z(nT)\sigma_i^z\ket{\psi_0}.
 \label{eq:app_local_autocorrelation}
\end{equation}
Because the initial product state satisfies
\begin{equation}
 \sigma_i^z\ket{\psi_0}=m_i^{(0)}\ket{\psi_0},
 \qquad
 m_i^{(0)}=\bra{\psi_0}\sigma_i^z\ket{\psi_0}=\pm1,
 \label{eq:app_initial_spin_pattern}
\end{equation}
we obtain
\begin{align}
 R_i(nT)
 &=m_i^{(0)}
 \bra{\psi_0}\left(U_F^\dagger\right)^n
 \sigma_i^zU_F^n\ket{\psi_0}
 \nonumber\\
 &=m_i^{(0)}\bra{\psi_n}\sigma_i^z\ket{\psi_n}.
 \label{eq:app_autocorrelation_identity_local}
\end{align}
Consequently,
\begin{align}
 \frac1L\sum_{i=1}^{L}
 \bra{\psi_0}\sigma_i^z(nT)\sigma_i^z\ket{\psi_0}
 &=\frac1L\sum_{i=1}^{L}m_i^{(0)}
 \bra{\psi_n}\sigma_i^z\ket{\psi_n}
 \nonumber\\
 &=R(nT).
 \label{eq:app_autocorrelation_identity}
\end{align}
Equation~\eqref{eq:app_autocorrelation_identity} relies on
$\ket{\psi_0}$ being an eigenstate of every local $\sigma_i^z$.  It does
not hold in this form for a general coherent or entangled initial state.

\section{Numerical controls and uncertainty analysis}
\label{app:mt_result_controls}

This appendix documents the data scope and the statistical,
observation-window, and finite-size controls underlying
Secs.~\ref{sec:rescaled_mt} and \ref{sec:mt_results}.  It also reports the decomposition and tightness tests
used in Sec.~\ref{sec:mt_results}.

\subsection{Data scope and bootstrap construction}
\label{app:numerical_data_scope}

The dynamical data for $L=8,10,12$ contain the stroboscopic spin response,
endpoint Fubini–Study angle, and path-averaged energy dispersion for each
of $N_{\rm dis}=50$ disorder realizations on the $39\times39$ grid.
The grid contains uniformly spaced values of
$J_zT/\hbar\in[0,1]$ and $\epsilon\in[0,0.4]$, and the trajectories extend
to $N=100$ Floquet periods.

The $L=16$ data contain realization-resolved endpoint angles
and path-averaged dispersions only at $n=100$, on a native $40\times40$ grid
of uniformly spaced values over the same parameter intervals. Consequently,
$L=16$ enters the fixed-endpoint normalization and coefficient analysis but
not the time-dependent, locked-response, or parity calculations. For the
four-size endpoint fields in Figs.~\ref{fig:mt_scaling_validation_revised}
and \ref{fig:app_mt_scaling_bootstrap}, the realization-resolved
$L=8,10,12$ records are bilinearly aligned from their native $39\times39$
axes to the native $L=16$ axes. The aligned records are bilinear interpolants
of the native data.

The quasienergy calculations use an independently generated
disorder ensemble on the nested $20\times20$ grid formed by every second
coordinate of the $39\times39$ dynamical axes. Realization-resolved
adjacent-gap data are available for $L=8$ and $L=10$, which are the only sizes
used in the spectral comparison. The response and spectral fields and their
crossover contours are evaluated on their native grids without interpolation. No
smoothing is applied to the four-size endpoint fields beyond the stated
cross-grid alignment.

The three fixed $L=10$ cells used for the trajectory illustrations are
$(J_zT/\hbar,\epsilon)=(0.789,0.042)$, $(0.211,0.295)$, and
$(0.789,0.358)$.  Their DTC-like, localized non-DTC, and thermal-like labels,
respectively, are assigned from the response and spectral diagnostics without
using the MT diagnostic.

At $L=10$, cells with $[r]_{\rm dis}>r_{\rm th}$ are labeled thermal-like.
Within the localized-like region $[r]_{\rm dis}<r_{\rm th}$, the
low-pulse-error side of the response crossover that is continuously connected to
the large-$h_\pi$ response is labeled DTC-like, and the opposite side is
labeled localized non-DTC.  These are operational finite-size and finite-time
assignments; the response and spectral constructions are given in
Appendices~\ref{app:response_crossover_uncertainty} and
\ref{app:mt_spectral_comparison}.

Within each size, every bootstrap draw uses the same resampled vector of
realization indices across the full parameter grid.  This preserves the
correlations induced by reusing each disorder realization
across the grid.  Disorder realizations are not
paired across system sizes, so the size ensembles are resampled
independently.  Dynamical MT and response quantities are resampled jointly
because they were evaluated on the same trajectories.  The spectral ensemble
is resampled separately from the dynamical data.  Every bootstrap calculation
uses 2000 nonparametric resamples, with the fixed seed 20260720.
To obtain the
bootstrap interval for the grid median of the relative
angular correction, we hold fixed the
angle-selected subset defined from the original
sample.  In each draw, $\mathcal T_{100,16}$ and
$C^{\rm sat}_{100,16}$ are recomputed using the same resampled vector of
$L=16$ realization indices before the median is taken over the selected
cells.

\subsection{Bootstrap uncertainty of the size normalization}
\label{app:mt_scaling_bootstrap}

For each bootstrap draw and parameter cell, let
$x_L=\ln[\overline{\Delta E}_{100,L}]_{\rm dis}$,
$\ell_L=\ln L$, and $\mathcal S=\{8,10,12,16\}$.  We recompute the
four-size logarithmic slope
\begin{equation}
 \gamma_{\mathcal S}
 =\frac{\sum_{L\in\mathcal S}
 (\ell_L-\bar\ell)(x_L-\bar x)}
 {\sum_{L\in\mathcal S}(\ell_L-\bar\ell)^2},
 \label{eq:app_gamma_estimator}
\end{equation}
where the bars in Eq.~\eqref{eq:app_gamma_estimator} denote averages
over the four sizes, not time or disorder averages.  We likewise recompute
the relative size range after defining
$s_L=[\overline{\Delta E}_{100,L}/\sqrt L]_{\rm dis}$
\begin{equation}
 \delta_{\mathcal S}\sigma
 =\frac{\max_{L\in\mathcal S}s_L-\min_{L\in\mathcal S}s_L}
 {\operatorname{mean}_{L\in\mathcal S}s_L}.
 \label{eq:app_sigma_range_estimator}
\end{equation}

To test the sensitivity of a candidate large-$L$ coefficient to the assumed
finite-size correction, we fit
\begin{equation}
 \mathcal T_{100,L}=C_p+a_pL^{-p},
 \qquad p\in\{1/2,1,2\},
 \label{eq:app_correction_models}
\end{equation}
and report the relative intercept spread
\begin{equation}
 \delta C
 =\frac{\max_p C_p-\min_p C_p}
 {\left|\operatorname{median}_p C_p\right|}.
 \label{eq:app_intercept_spread}
\end{equation}
This quantity measures sensitivity to the assumed correction
form and is not a statistical error bar.

\begin{figure*}[t] \centering \includegraphics[height=0.37\textheight,width=0.95\textwidth] {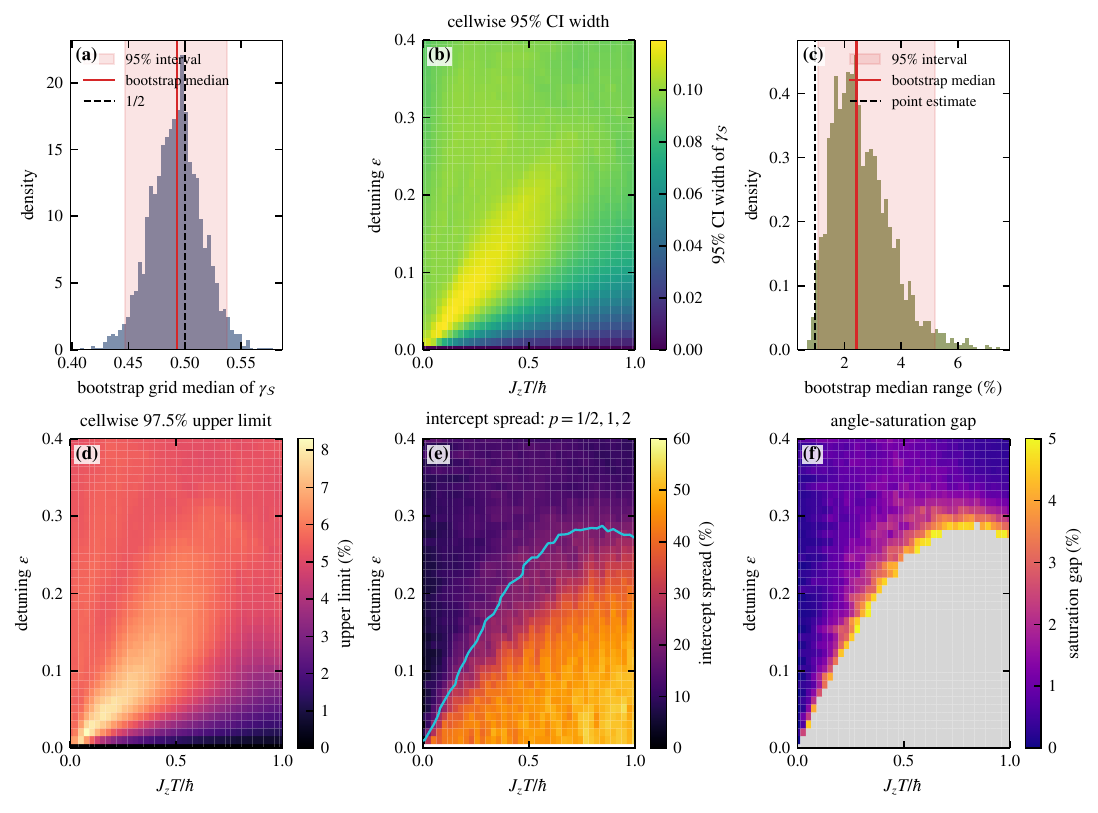}
 \caption{Four-size statistical controls and sensitivity
 to the assumed correction form at
 $n=100$ on the native $40\times40$ $L=16$ parameter grid.
 The realization-resolved $L=8,10,12$ endpoint records are bilinearly aligned
 from their native $39\times39$ axes to the $L=16$ axes; the aligned values
 are bilinear interpolants of the native records. Each native or aligned cell contains
 $N_{\rm dis}=50$ realizations; all bootstrap results use 2000
 resamples.  (a) Bootstrap distribution of the median
 exponent over the grid, obtained by independent resampling at each size;
 the red line marks the
 bootstrap median, the shaded red band shows
 its 95\% interval, and the dashed line marks $1/2$.
 (b) Cellwise width of the exponent's 95\% bootstrap interval.
 (c) Bootstrap distribution of the grid median of
 Eq.~\eqref{eq:app_sigma_range_estimator}; the dashed line is the point
 estimate.  (d) Cellwise 97.5th-percentile upper limit on the
 range of the normalized dispersion.  (e) Relative intercept spread from
 Eq.~\eqref{eq:app_intercept_spread}; values above $60\%$ share the upper
 color and the cyan contour marks
 $[\mathcal L_{100,16}]_{\rm dis}=0.95(\pi/2)$.  (f) Exact
 relative angular correction
 $100(1-\mathcal T_{100,16}/C^{\rm sat}_{100,16})$ within the
 angle-selected cells; other cells are gray.  Sizes are resampled
 independently, while one realization-index vector is shared over the
 parameter grid at fixed size. No smoothing is applied beyond
 the stated cross-grid alignment.}
 \label{fig:app_mt_scaling_bootstrap}
\end{figure*}

The point estimate of the median exponent over the grid is
$0.492$.  Its bootstrap
median is $0.493$, and its 95\% percentile interval is
$0.447$–$0.537$
[Fig.~\ref{fig:app_mt_scaling_bootstrap}(a)].  The median and 95th percentile
of the cellwise interval widths are $0.096$ and
$0.111$, respectively
[Fig.~\ref{fig:app_mt_scaling_bootstrap}(b)].
{Within numerical tolerance, all cellwise intervals contain $1/2$;
for the 40 cells on $\epsilon=0$, the lower limits exceed $1/2$ by less than
$3\times10^{-9}$.}
The cellwise interval widths quantify the statistical precision of the
four-size exponent test, which remains a finite-size consistency check rather
than a precise asymptotic estimate.

{The point estimate of the grid median of
$\delta_{\mathcal S}\sigma$ is $1.0\%$.}
{The corresponding bootstrap distribution has median $2.4\%$ and a
95\% interval of $1.1\%$–$5.2\%$
[Fig.~\ref{fig:app_mt_scaling_bootstrap}(c)].}
The displacement of the
bootstrap median from the point estimate is expected for a nonnegative
max–min statistic: independent sampling fluctuations at the four sizes
increase the range even when the underlying means are close.
{The cellwise 97.5th-percentile upper limits reach at most $8.3\%$ on
the sampled grid [Fig.~\ref{fig:app_mt_scaling_bootstrap}(d)].}
The bootstrap test therefore
supports $\sqrt L$ as a leading normalization at the current precision; it
does not establish sub-percent finite-size collapse.

Sensitivity to the assumed correction form is the dominant
limitation on a pointwise
extrapolation of the full size-rescaled MT diagnostic.
{Across cells with $\epsilon>0$, the median spread of the three fitted intercepts is
$20.1\%$, and its 95th percentile across parameter cells is $49.8\%$.}
{Within the angle-selected subset, the corresponding values are
$12.8\%$ and $19.5\%$
[Fig.~\ref{fig:app_mt_scaling_bootstrap}(e)].}
{By comparison, the exact relative angular correction in this subset
has a median of $1.0\%$ and a 95th percentile of $3.8\%$ across parameter
cells [Fig.~\ref{fig:app_mt_scaling_bootstrap}(f)].}
The angle criterion selects 795 of the 1560 cells with
$\epsilon>0$.
This fixed-subset bootstrap gives a 95\% percentile interval
of $0.92\%$–$1.08\%$ for the grid median of the
relative angular correction.
{Within the same fixed subset, the raw point
estimate of the median symmetric difference
$d_{12,16}(C^{\rm sat}_{100,L})$ is $0.2\%$.}
{The bootstrap distribution of this median symmetric relative
difference has median $0.9\%$ and a central 95\% interval of
$0.2\%$–$3.2\%$; the corresponding median signed difference between $L=12$
and $L=16$ is compatible with zero.}
These
results are consistent with a finite coefficient in the selected subset, but
the dependence on the assumed correction form precludes a controlled pointwise
extrapolation.

\subsection{Dependence of the normalization test on endpoint time}
\label{app:mt_time_normalization}

\begin{figure*}[t]
 \centering
 \includegraphics[width=\textwidth]
 {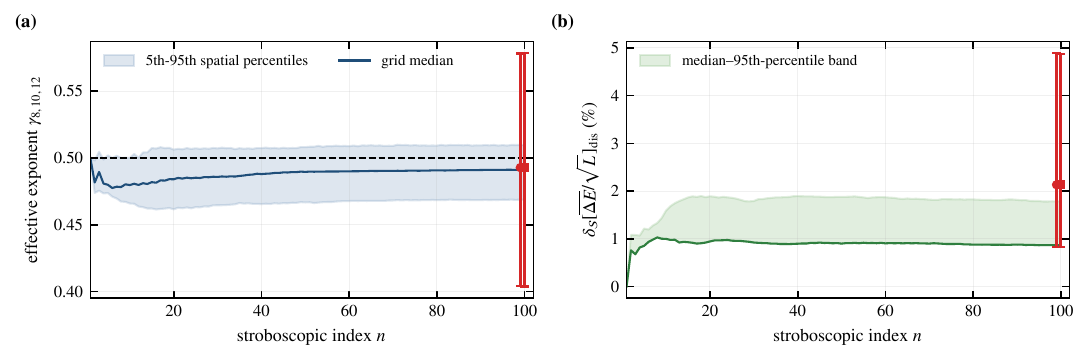}
 \caption{Time-dependent normalization controls for $L=8,10,12$ on the
 $39\times39$ grid over $1\leq n\leq100$.
 (a) Grid median and 5th–95th percentile range of the
 effective exponent across parameter cells;
 red symbols show 95\% bootstrap intervals obtained
 by independent resampling at each size at
 $n=99,100$.  (b) Grid median and 95th percentile of the
 relative range of the normalized dispersion across parameter cells, with
 95\% bootstrap intervals for the median at
 the same two endpoints.  Panel (b) reports percentages;  $N_{\rm dis}=50$ and 2000 bootstrap resamples are used.}
 \label{fig:app_mt_time_normalization}
\end{figure*}

Because the $L=16$ record contains only the $n=100$ endpoint, the following
complementary time-dependent test uses the $39\times39$ trajectory data for
$L=8,10,12$.  The three-size test is
repeated at every stroboscopic endpoint
$1\leq n\leq100$.
{At each $n$, the dark curve in
Fig.~\ref{fig:app_mt_time_normalization}(a) is the median over all parameter
cells, and the shaded region is their 5th–95th spatial percentile interval.}
The statistical bootstrap uses the realization-level
angle and dispersion arrays retained at $n=99$ and $100$; the red intervals show the resulting uncertainty in the grid median.  The shaded region measures variation over $(J_zT/\hbar,\epsilon)$, and the red intervals quantify uncertainty from the finite disorder sample.

At $n=1$, the dispersion is generated solely by the transverse segment and
obeys $\overline{\Delta E}_{1,L}\propto\sqrt L$ exactly, giving
$\gamma_{8,10,12}=1/2$ and zero range of the normalized
dispersion to numerical
precision.  The point estimate of the median exponent over
the grid
lies between $0.477$ and $0.480$ over $n=5$–$10$ and is $0.491$ at $n=100$.  The
bootstrap intervals at $n=99$ and $100$ contain $1/2$.  Over the same
trajectory, the point estimate of the grid median of the
range of the normalized dispersion
remains near $1\%$ after the initial transient, and its late-time
95th percentile across parameter cells is near $1.8\%$
[Fig.~\ref{fig:app_mt_time_normalization}(b)].

Thus the denominator is compatible with the same leading normalization
throughout the sampled trajectory.  The endpoint-angle drift and parity
dependence documented in Figs.~\ref{fig:mt_scaling_validation_revised} and
\ref{fig:mt_endpoint_parity_revised} nevertheless preclude a
time-independent size collapse of the full size-rescaled MT diagnostic.

\subsection{Sampling and window dependence of the response
crossover}
\label{app:response_crossover_uncertainty}

For a stroboscopic window $[a,b]$, the locked spin response of
realization $s$ is
\begin{equation}
 h_{\pi,s,L}^{[a,b]}
 =\left|
 \frac{1}{b-a+1}\sum_{n=a}^{b}(-1)^n
 R_{s,L}(nT)
 \right|.
 \label{eq:app_locked_response_window}
\end{equation}
For an ideal period-doubled trajectory
$R_{s,L}(nT)=(-1)^n$,
Eq.~\eqref{eq:app_locked_response_window} gives
$h_{\pi,s,L}^{[a,b]}=1$.

The response-crossover estimator at fixed $J_zT/\hbar$ is the sampled
pulse error
\begin{equation}
 \epsilon_{{\rm resp},L}^{*}(J_zT/\hbar;[a,b])
 =\underset{\epsilon\ \mathrm{on\ the\ grid}}{\arg\max}\;
 \operatorname{Var}_{\rm dis}
 \left(h_{\pi,s,L}^{[a,b]}\right).
 \label{eq:app_response_crossover_estimator}
\end{equation}
The maximization is performed directly over the sampled values of
$\epsilon$, without interpolation or a monotonicity constraint.  The
central estimator uses $[a,b]=[1,100]$.  Its disorder-sampling uncertainty is
represented by the 16th–84th percentile interval of the bootstrap
distribution of
$\epsilon_{{\rm resp},L}^{*}
(J_zT/\hbar;[1,100])$.

\begin{figure*}[t]
 \centering
 \includegraphics[width=\textwidth]
 {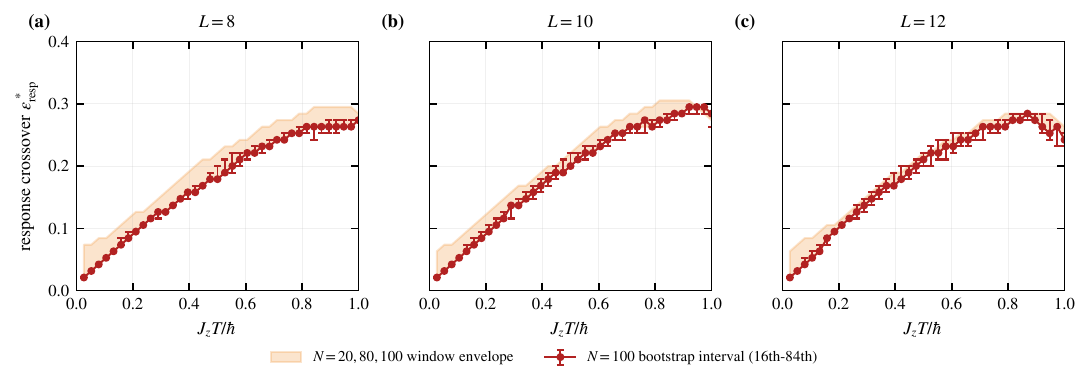}
 \caption{Disorder-sampling, observation-window, and finite-size controls for
 the response-crossover estimator.  Red markers show
 $\epsilon_{{\rm resp},L}^{*}
 (J_zT/\hbar;[1,100])$ for $L=8,10,12$, and the
 red error bars denote the 16th–84th percentile intervals from 2000
 bootstrap resamples of the disorder realizations.  The orange
 band is the envelope obtained
 from the windows $[1,N]$ with $N=20,80,100$.  Each parameter cell contains
 $N_{\rm dis}=50$ realizations.  All estimates remain on the sampled
 $39\times39$ grid, whose pulse-error spacing is
 $\Delta\epsilon=0.4/38\simeq0.0105$.}
 \label{fig:app_response_uncertainty}
\end{figure*}

Observation-window dependence is assessed separately by repeating
Eq.~\eqref{eq:app_response_crossover_estimator} for $[a,b]=[1,N]$ with
$N=20,80,100$.  The resulting envelope, shown together with the
disorder-bootstrap and finite-size controls in
Fig.~\ref{fig:app_response_uncertainty}, measures finite-time sensitivity
and is distinct from the bootstrap interval over disorder
realizations.

{For the central $[1,100]$ estimator, the cross-size range of the
$L=8,10,12$ crossover coordinates has a median width of two and a maximum
width of four pulse-error grid cells over the $J_zT/\hbar>0$ columns.}
For $L=8,10,12$, the median (maximum) full widths
of the observation-window envelopes are three (five), two (four), and one
(four) cells, respectively.  The corresponding median (maximum) widths of
the 16th–84th percentile bootstrap intervals are one (three), one (two),
and two (three) cells.

Finite disorder sampling and observation-window variation therefore produce
resolvable variation in the estimator on the scale of the available
parameter grid.  These controls quantify the stability of the operational
response crossover over the available sizes and observation windows.  A
thermodynamic DTC boundary requires a dedicated finite-size scaling analysis.

\subsection{Finite-size spectral comparison}
\label{app:mt_spectral_comparison}

This subsection tests the finite-size drift and disorder-sampling sensitivity
of the operational adjacent-gap crossover used in the main text.

{
For each size $L$ and disorder realization $s$, let the $D=2^L$
eigenphases of $U_F$ be ordered as
\begin{equation}
 0\leq\theta_{1,L}^{(s)}
 \leq\theta_{2,L}^{(s)}\leq\cdots
 \leq\theta_{D,L}^{(s)}<2\pi.
\end{equation}
The eigenphase spacings on the Floquet circle are
\begin{align}
 \delta_{k,L}^{(s)}
 &=\theta_{k+1,L}^{(s)}-\theta_{k,L}^{(s)},
 \qquad 1\leq k<D,
 \nonumber\\
 \delta_{D,L}^{(s)}
 &=2\pi+\theta_{1,L}^{(s)}-\theta_{D,L}^{(s)}.
 \label{eq:app_floquet_spacings}
\end{align}
With the spacing index understood cyclically, the adjacent-gap ratio is
\begin{equation}
 r_{k,L}^{(s)}=
 \frac{\min(\delta_{k,L}^{(s)},\delta_{k+1,L}^{(s)})}
 {\max(\delta_{k,L}^{(s)},\delta_{k+1,L}^{(s)})}.
 \label{eq:app_gap_ratio}
\end{equation}}

Ratios involving a numerically unresolved spacing are omitted.  More
precisely, for realization $s$ and size $L$ we define
\begin{equation}
 {
 \mathcal K_{s,L}=
 \left\{k:\delta_{k,L}^{(s)}>10^{-12},\
 \delta_{k+1,L}^{(s)}>10^{-12}\right\},}
 \label{eq:app_valid_gap_set}
\end{equation}
and calculate
\begin{equation}
 {
 \overline r_{s,L}
 =\frac{1}{|\mathcal K_{s,L}|}
 \sum_{k\in\mathcal K_{s,L}}r_{k,L}^{(s)}.}
 \label{eq:app_realization_gap_average}
\end{equation}
The spectral field used in the crossover analysis is the subsequent
disorder mean
\begin{equation}
 {
 [r_L]_{\rm dis}
 =\frac{1}{N_{\rm dis}}\sum_{s=1}^{N_{\rm dis}}\overline r_{s,L}.}
 \label{eq:app_disorder_gap_average}
\end{equation}
At $(J_zT/\hbar,\epsilon)=(0,0)$, exact degeneracies leave no valid
adjacent ratios, and the corresponding cell is excluded from the spectral
map.

The operational threshold used throughout is
$r_{\rm th}\equiv(0.386+0.527)/2=0.4565$, where
$r_{\rm P}=2\ln 2-1\simeq0.386$ and
$r_{\rm COE}\simeq0.527$ are the Poisson and
circular-orthogonal-ensemble reference values, respectively
\cite{OganesyanHuse2007,Atas2013,DAlessioRigol2014}.
The bootstrap probability
\begin{equation}
 p_L(J_zT/\hbar,\epsilon)
 =P_{\rm boot}\!\left([r_L]_{\rm dis}>r_{\rm th}\right),
 \label{eq:app_spectral_bootstrap_probability}
\end{equation}
quantifies the sensitivity of the finite-size spectral
classification to the available disorder ensemble; it is neither a phase
probability nor a determination of a thermodynamic localization boundary.

\begin{figure*}[t]
 \centering
 \includegraphics[width=\textwidth]
 {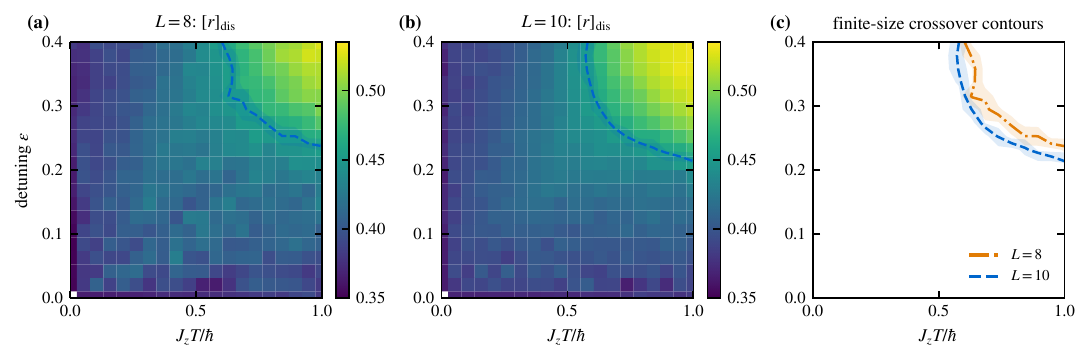}
 \caption{Finite-size spectral comparison from realization-resolved data with
 $N_{\rm dis}=50$ and 2000 bootstrap
 resamples.  (a) and (b) show $[r]_{\rm dis}$ for $L=8$ and $L=10$,
 respectively, on the common range
 $0.35\leq[r]_{\rm dis}\leq0.535$.  Solid contours mark
 $[r]_{\rm dis}=r_{\rm th}=0.4565$; shaded regions identify
 $0.16\leq p_L\leq0.84$, where the threshold classification is
 bootstrap-sensitive.  (c) compares the $L=8$ and $L=10$ contours and
 their bootstrap-sensitive regions.  The cell
 $(J_zT/\hbar,\epsilon)=(0,0)$ is blank because its gap ratio is undefined.
 All panels display the native $20\times20$ cells without interpolation or
 smoothing.}
 \label{fig:app_mt_spectral_comparison}
\end{figure*}

Both sizes resolve a region with comparatively large $[r]_{\rm dis}$ at
simultaneously large $J_zT/\hbar$ and $\epsilon$
[Figs.~\ref{fig:app_mt_spectral_comparison}(a) and
\ref{fig:app_mt_spectral_comparison}(b)].  The mean threshold contours show
a finite displacement between $L=8$ and $L=10$.  The bootstrap-sensitive
set $0.16\leq p_L\leq0.84$ contains nine cells for $L=8$ and seven cells
for $L=10$, with partial overlap between the two sizes.  The $L=10$ contour
is therefore used as the spectral crossover at the largest
accessible size in the
main-text comparison, while the $L=8$–$L=10$ displacement provides a
control on finite-size drift.

\subsection{Decomposition and tightness of the MT lower-bound functional}
\label{app:mt_time_components}

\begin{figure*}[t]
 \centering
 \includegraphics[width=\textwidth]
 {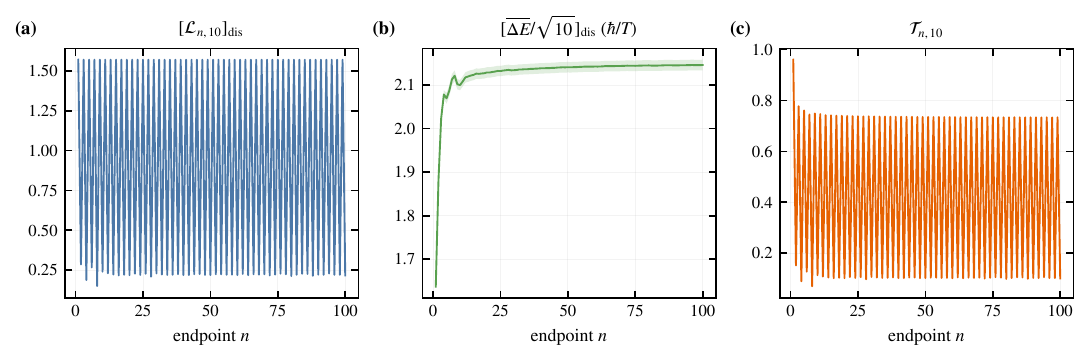}
 \caption{{Time-resolved MT components at the representative DTC-like
 $L=10$ cell $(J_zT/\hbar,\epsilon)=(0.789,0.042)$.
 Panels (a)–(c) show $[\mathcal L_{n,10}]_{\rm dis}$,
 $[\sigma_{n,10}]_{\rm dis}=
 [\overline{\Delta E}_{n,10}/\sqrt{10}]_{\rm dis}$, and
 $\mathcal T_{n,10}$, respectively.  Curves are disorder means and bands are
 standard errors for $N_{\rm dis}=50$.}}
 \label{fig:app_mt_time_components}
\end{figure*}

The odd–even structure of the size-rescaled MT diagnostic can arise from
the endpoint angle, the temporal modulation of the normalized dispersion in
the denominator, or both.  The time-resolved quantities used in this
comparison and in the tightness test are
\begin{equation}
 \begin{aligned}
  &\mathcal L_{n,10}^{(s)},\qquad
  \sigma_{n,10}^{(s)}
  =\frac{\overline{\Delta E}_{n,10}^{(s)}}{\sqrt{10}},\\
  &\mathcal T_{n,10},\qquad
  q_{n,10}^{(s)}
  =\frac{\tau_{{\rm MT},n,10}^{(s)}}{nT}.
 \end{aligned}
 \label{eq:app_mt_component_definitions}
\end{equation}
{
Figure~\ref{fig:app_mt_time_components} shows the disorder-averaged endpoint
angle, normalized dispersion, and rescaled MT diagnostic at the
 representative DTC-like cell defined in
Sec.~\ref{subsec:mt_endpoint_parity}.  The tightness ratio
$q_{n,10}^{(s)}$ is evaluated separately for each
realization and satisfies $0\leq q_{n,10}^{(s)}\leq1$.}

For a window $[a,b]$, let $M=b-a+1$ and define
\begin{equation}
 \bar\sigma_{L,s}^{[a,b]}
 =
 \frac{1}{M}\sum_{n=a}^{b}\sigma_{n,L}^{(s)}.
 \label{eq:app_frozen_sigma}
\end{equation}
The signed coefficient whose absolute value enters
Eq.~\eqref{eq:results_locked_mt} admits the exact decomposition
\begin{align}
 A_{\pi,L,s}^{\rm MT,[a,b]}
 &:=\frac{1}{M}\sum_{n=a}^{b}(-1)^n
 \frac{\hbar\mathcal L_{n,L}^{(s)}}{T\sigma_{n,L}^{(s)}}
 =A_{\pi,L,s}^{\rm fr,[a,b]}+A_{\pi,L,s}^{\rm den,[a,b]},
 \label{eq:app_mt_frozen_decomposition}\\
 A_{\pi,L,s}^{\rm fr,[a,b]}
 &:=\frac{\hbar}{T\bar\sigma_{L,s}^{[a,b]}}
 \frac{1}{M}\sum_{n=a}^{b}(-1)^n\mathcal L_{n,L}^{(s)},
 \nonumber\\
 A_{\pi,L,s}^{\rm den,[a,b]}
 &:=\frac{\hbar}{TM}\sum_{n=a}^{b}(-1)^n\mathcal L_{n,L}^{(s)}
 \left[
 \frac{1}{\sigma_{n,L}^{(s)}}-\frac{1}{\bar\sigma_{L,s}^{[a,b]}}
 \right].
 \nonumber
\end{align}
The fixed-denominator term retains the
realization-dependent dispersion scale while
removing its temporal modulation.  The identity in
Eq.~\eqref{eq:app_mt_frozen_decomposition} holds before the absolute value
and disorder average; consequently, the two disorder-averaged absolute
amplitudes are not themselves additive.  We use
``fixed-denominator amplitude'' and
``denominator-modulation amplitude'' to denote
$[|A_{\pi,L,s}^{\rm fr,[a,b]}|]_{\rm dis}$ and
$[|A_{\pi,L,s}^{\rm den,[a,b]}|]_{\rm dis}$, respectively.

{
{At the representative DTC-like cell, the fixed-denominator amplitude
nearly coincides with the full-window amplitude, while the
denominator-modulation amplitude is about $0.4\%$ of the full-window value.}
{Because the absolute value precedes disorder averaging, additivity
applies to the signed realization-level coefficients in
Eq.~\eqref{eq:app_mt_frozen_decomposition}, while the separately averaged
absolute amplitudes are reported individually.}}

{Across the 1482 cells with $J_zT/\hbar>0$, the full and
fixed-denominator maps have Spearman coefficient $0.998$.}
{Freezing the temporal modulation of the denominator therefore leaves
the ordering of the parameter cells nearly unchanged.}
{The normalized dispersion remains essential to the scale and
interpretation of the MT diagnostic.}
{At the representative DTC-like cell, its temporal modulation
contributes only weakly, and the locked MT component is carried predominantly
by the endpoint angle.}
The localized non-DTC and thermal-like cells show no comparable odd–even
branch separation [Fig.~\ref{fig:mt_endpoint_parity_revised}].

{At $n=100$, the realization-level tightness satisfies
$q_{100,L}^{(s)}<2.1\times10^{-3}$ for every available parameter cell and
disorder realization at $L=8,10,12,16$.}
{Thus the MT inequality is far from saturation at the endpoint used
for comparison across the parameter plane.}
{The reported MT structure characterizes the endpoint geometry and
path-averaged energy dispersion of trajectories in this far-from-saturation
regime.}

\section{Analytic controls for the MT size normalization}
\label{app:mt_normalization_controls}

This appendix establishes the assumptions behind the $\sqrt L$
normalization and derives the ideal-pulse sequence exactly.  All analytic
statements hold for each realization.  Disorder averaging
is performed only after the
physical MT lower-bound functional has been constructed for each
realization.  Statistical and finite-time controls are collected separately in
Appendix~\ref{app:mt_result_controls}.

\subsection{Covariance criterion for
\texorpdfstring{$\sqrt L$}{sqrt(L)} normalization}
\label{app:mt_covariance}

On segment $\alpha$, let
\begin{equation}
 H_{\alpha,L}=\sum_{x=1}^{M_{\alpha,L}}
 \mathfrak h_{\alpha,x},
 \qquad
 M_{\alpha,L}\leq\kappa_\alpha L+O(1),
 \label{eq:app_local_decomposition}
\end{equation}
where the $\mathfrak h_{\alpha,x}$ are Hermitian local
operators with norms bounded
uniformly in $L$, $x$, and the disorder realization.  In the evolving
normalized state, define
\begin{equation}
 \delta\mathfrak h_{\alpha,x}^{(s)}(t)
 =\mathfrak h_{\alpha,x}
 -\langle\mathfrak h_{\alpha,x}\rangle_{s,t},
 \label{eq:app_local_fluctuation}
\end{equation}
and the real, symmetric covariance matrix
\begin{equation}
 C_{\alpha,xy}^{(s)}(t)
 =\frac12\left\langle
 \{\delta\mathfrak h_{\alpha,x}^{(s)}(t),
   \delta\mathfrak h_{\alpha,y}^{(s)}(t)\}
 \right\rangle_{s,t}.
 \label{eq:app_covariance_definition}
\end{equation}
The Hamiltonian variance satisfies
\begin{align}
 \Delta^2H_{\alpha,L}^{(s)}(t)
 &=\sum_{x,y}
 \langle\delta\mathfrak h_{\alpha,x}^{(s)}(t)
 \delta\mathfrak h_{\alpha,y}^{(s)}(t)\rangle_{s,t}
 \nonumber\\
 &=\sum_{x,y}C_{\alpha,xy}^{(s)}(t).
 \label{eq:app_covariance_identity}
\end{align}
The second equality remains valid when distinct local terms do not commute
the double sum of the antisymmetric commutator contribution vanishes under
$x\leftrightarrow y$.

Assume that the connected covariances are absolutely and uniformly
summable over all samples, segments, sizes, and times included in the
observation window,
\begin{equation}
 \sup_{L,s,\alpha,t,x}
 \sum_y|C_{\alpha,xy}^{(s)}(t)|\leq C_*<\infty.
 \label{eq:app_covariance_uniform_bound}
\end{equation}
Equations~\eqref{eq:app_local_decomposition} and
\eqref{eq:app_covariance_uniform_bound} imply
\begin{align}
 0\leq\Delta^2H_{\alpha,L}^{(s)}(t)
 &\leq\sum_x\sum_y|C_{\alpha,xy}^{(s)}(t)|
 \nonumber\\
 &\leq M_{\alpha,L}C_*=O(L).
 \label{eq:app_variance_extensive_bound}
\end{align}
Thus $\Delta H_{\alpha,L}^{(s)}(t)=O(\sqrt L)$ uniformly.

For an equal-duration binary drive, the path-averaged dispersion is
\begin{equation}
 \overline{\Delta E}_{n,L}^{(s)}
 =\frac{1}{2n}\sum_{j=0}^{n-1}
 \left[
 \Delta_{\psi_{j,A}^{(s)}}(H_1)
 +\Delta_{\psi_{j,B}^{(s)}}(H_2)
 \right].
 \label{eq:app_binary_path_dispersion}
\end{equation}
The uniform bound therefore gives
$\overline{\Delta E}_{n,L}^{(s)}=O(\sqrt L)$ for every fixed $n$.
If the observation time grows with $L$, the same conclusion requires
Eq.~\eqref{eq:app_covariance_uniform_bound} to remain valid over the growing
time interval.  Equation~\eqref{eq:app_binary_path_dispersion} averages
instantaneous standard deviations.  The square root of a time-averaged
variance defines a different quantity.

The covariance argument supplies only
an $O(\sqrt L)$ upper bound.  The statement
$\overline{\Delta E}_{n,L}=\Theta(\sqrt L)$ requires
\begin{equation}
 0<\liminf_{L\rightarrow\infty}
 \frac{\overline{\Delta E}_{n,L}}{\sqrt L}
 \leq
 \limsup_{L\rightarrow\infty}
 \frac{\overline{\Delta E}_{n,L}}{\sqrt L}<\infty.
 \label{eq:app_nonzero_sigma_condition}
\end{equation}
Nonsummable long-range or critical connected
correlations \cite{Heyl2017}, macroscopic
cat states \cite{Frowis2012}, or a vanishing variance density can invalidate this stronger
condition.  Even when Eq.~\eqref{eq:app_nonzero_sigma_condition} holds, the
size-rescaled MT quantity for a given realization can vanish if its endpoint
angle vanishes.  The separate dispersion and angle data therefore determine
whether the $\sqrt L$ normalization applies.

For the N\'eel initial state, which is a $\sigma^z$ product state,
$\Delta_{\psi_0}(H_1)=0$ and
$U_1|\psi_0\rangle=e^{i\phi_s}|\psi_0\rangle$ for each disorder realization.
At the beginning of the first transverse segment,
$\langle H_2\rangle=0$ and
$\langle H_2^2\rangle
=\Omega(\epsilon)^2\sum_{i,j}\langle\sigma_i^x\sigma_j^x\rangle
=\Omega(\epsilon)^2L$, because
$\langle\sigma_i^x\sigma_j^x\rangle=0$ for $i\neq j$ in a
$\sigma^z$ product state. Hence
$\Delta_{\psi_{0,B}^{(s)}}(H_2)
=|\Omega(\epsilon)|\sqrt L$.
All remaining terms in Eq.~\eqref{eq:app_binary_path_dispersion} are
nonnegative, so
$\overline{\Delta E}_{n,L}^{(s)}
\geq|\Omega(\epsilon)|\sqrt L/(2n)$.
Division by $\sqrt L$ gives
$\sigma_{n,L}^{(s)}\geq|\Omega(\epsilon)|/(2n)$. This lower bound is
independent of $L$ at fixed $n$, but decreases as $1/n$ and is therefore not
uniform when the observation time grows with system size. This result is a
lower bound on $\sigma_{n,L}^{(s)}$; the covariance summability condition and
endpoint-angle behavior remain separate requirements.

\subsection{Exact ideal-pulse recurrence}
\label{app:mt_ideal_pulse}

For the disordered Ising chain with open boundary conditions, decompose
\begin{equation}
 \begin{aligned}
 H_1&=H_J+H_h,\\
 H_J&=\sum_{i=1}^{L-1}J_{z,i}\sigma_i^z\sigma_{i+1}^z,
 &H_h&=\sum_{i=1}^{L}h_i\sigma_i^z,
 \end{aligned}
 \label{eq:app_HJ_Hh}
\end{equation}
and let $P=\prod_{i=1}^L\sigma_i^x$.  With segment duration $T/2$, the ideal
pulse condition is
\begin{equation}
 \frac{\Omega T}{2\hbar}=\frac{\pi}{2},
 \qquad
 U_2=e^{-i(\pi/2)\sum_i\sigma_i^x}=(-i)^LP.
 \label{eq:app_ideal_U2}
\end{equation}
Global spin flip leaves the interaction invariant and reverses the
longitudinal field,
\begin{equation}
 PH_JP=H_J,
 \qquad
 PH_hP=-H_h,
 \qquad
 [H_J,H_h]=0.
 \label{eq:app_P_conjugation}
\end{equation}
For $U_F=U_2U_1$, with
$U_1=e^{-i(H_J+H_h)T/(2\hbar)}$, it follows that
\begin{align}
 U_F^2
 &=(-1)^L
 e^{-i(H_J-H_h)T/(2\hbar)}
 e^{-i(H_J+H_h)T/(2\hbar)}
 \nonumber\\
 &=(-1)^L e^{-iH_JT/\hbar}.
 \label{eq:app_UF_squared}
\end{align}

Every $\sigma^z$ product state is an eigenstate of $H_J$.  For the N\'eel
state there are phases $\phi_m$ and $\chi_m$ such that
\begin{equation}
 U_F^{2m}|\psi_0\rangle=e^{i\phi_m}|\psi_0\rangle,
 \qquad
 U_F^{2m+1}|\psi_0\rangle=e^{i\chi_m}P|\psi_0\rangle.
 \label{eq:app_ideal_orbit}
\end{equation}
The states $|\psi_0\rangle$ and $P|\psi_0\rangle$ are orthogonal, giving
\begin{equation}
 \mathcal L_{2m,L}=0,
 \qquad
 \mathcal L_{2m+1,L}=\frac{\pi}{2}.
 \label{eq:app_ideal_angles}
\end{equation}

At the start of each segment, the state is a $\sigma^z$ product state up to
a phase.  Hence $\Delta H_1=0$.  On the transverse segment,
\begin{equation}
 \langle\sigma_i^x\rangle=0,
 \qquad
 \langle\sigma_i^x\sigma_j^x\rangle=0
 \quad(i\neq j),
 \label{eq:app_ideal_x_expectations}
\end{equation}
and therefore
\begin{equation}
 \Delta H_2=|\Omega|\sqrt L
 =\frac{\pi\hbar}{T}\sqrt L.
 \label{eq:app_ideal_H2_variance}
\end{equation}
Substitution into Eq.~\eqref{eq:app_binary_path_dispersion} yields, for
every $n\geq1$,
\begin{equation}
 \frac{\overline{\Delta E}_{n,L}}{\sqrt L}
 =\frac{\pi\hbar}{2T}.
 \label{eq:app_ideal_sigma}
\end{equation}
Combining Eqs.~\eqref{eq:app_ideal_angles} and
\eqref{eq:app_ideal_sigma} gives
\begin{equation}
 \frac{\widetilde\tau_{{\rm MT},n,L}}{T}
 =\begin{cases}
 1,&n\ \mathrm{odd},\\
 0,&n\ \mathrm{even}.
 \end{cases}.
 \label{eq:app_ideal_MT_result}
\end{equation}
The even-period zero is caused by exact return of the endpoint ray, not by a
vanishing path-averaged energy dispersion.
The derivation remains valid at $J_z=0$ and establishes the
exact normalization and implementation check stated in
Eq.~\eqref{eq:ideal_mt_sequence}.

\ifprbclass
  \bibliographystyle{apsrev4-2}
\else
  \bibliographystyle{unsrt}
\fi
\bibliography{references}

@article{MandelstamTamm1945,
  author  = {Mandelstam, Leonid and Tamm, Igor},
  title   = {The uncertainty relation between energy and time in non-relativistic quantum mechanics},
  journal = {J. Phys. (USSR)},
  volume  = {9},
  pages   = {249--254},
  year    = {1945}
}

@article{AnandanAharonov1990,
  author  = {Anandan, Jeeva and Aharonov, Yakir},
  title   = {Geometry of quantum evolution},
  journal = {Phys. Rev. Lett.},
  volume  = {65},
  pages   = {1697--1700},
  year    = {1990},
  doi     = {10.1103/PhysRevLett.65.1697}
}

@article{Giovannetti2003,
  author        = {Giovannetti, Vittorio and Lloyd, Seth and Maccone, Lorenzo},
  title         = {Quantum limits to dynamical evolution},
  journal       = {Phys. Rev. A},
  volume        = {67},
  pages         = {052109},
  year          = {2003},
  doi           = {10.1103/PhysRevA.67.052109}
}

@article{Frowis2012,
  author        = {Fr{\"o}wis, Florian},
  title         = {Kind of entanglement that speeds up quantum evolution},
  journal       = {Phys. Rev. A},
  volume        = {85},
  pages         = {052127},
  year          = {2012},
  doi           = {10.1103/PhysRevA.85.052127}
}

@article{DAlessioRigol2014,
  author        = {D'Alessio, Luca and Rigol, Marcos},
  title         = {Long-time behavior of isolated periodically driven interacting lattice systems},
  journal       = {Phys. Rev. X},
  volume        = {4},
  pages         = {041048},
  year          = {2014},
  doi           = {10.1103/PhysRevX.4.041048}
}

@article{Ponte2015,
  author        = {Ponte, Pedro and Papi{\'c}, Z. and Huveneers, Fran{\c c}ois and Abanin, Dmitry A.},
  title         = {Many-body localization in periodically driven systems},
  journal       = {Phys. Rev. Lett.},
  volume        = {114},
  pages         = {140401},
  year          = {2015},
  doi           = {10.1103/PhysRevLett.114.140401}
}

@article{Khemani2016,
  author        = {Khemani, Vedika and Lazarides, Achilleas and Moessner, Roderich and Sondhi, S. L.},
  title         = {Phase structure of driven quantum systems},
  journal       = {Phys. Rev. Lett.},
  volume        = {116},
  pages         = {250401},
  year          = {2016},
  doi           = {10.1103/PhysRevLett.116.250401}
}

@article{Else2016,
  author        = {Else, Dominic V. and Bauer, Bela and Nayak, Chetan},
  title         = {{Floquet} time crystals},
  journal       = {Phys. Rev. Lett.},
  volume        = {117},
  pages         = {090402},
  year          = {2016},
  doi           = {10.1103/PhysRevLett.117.090402}
}

@article{vonKeyserlingk2016,
  author        = {{von Keyserlingk}, Curt W. and Khemani, Vedika and Sondhi, S. L.},
  title         = {Absolute stability and spatiotemporal long-range order in {Floquet} systems},
  journal       = {Phys. Rev. B},
  volume        = {94},
  pages         = {085112},
  year          = {2016},
  doi           = {10.1103/PhysRevB.94.085112}
}

@article{Yao2017,
  author        = {Yao, Norman Y. and Potter, Andrew C. and Potirniche, Ionut-Dragos and Vishwanath, Ashvin},
  title         = {Discrete time crystals: Rigidity, criticality, and realizations},
  journal       = {Phys. Rev. Lett.},
  volume        = {118},
  pages         = {030401},
  year          = {2017},
  doi           = {10.1103/PhysRevLett.118.030401},
  note          = {Erratum: Phys. Rev. Lett. 118, 269901 (2017)}
}

@article{ElsePrethermal2017,
  author        = {Else, Dominic V. and Bauer, Bela and Nayak, Chetan},
  title         = {Prethermal phases of matter protected by time-translation symmetry},
  journal       = {Phys. Rev. X},
  volume        = {7},
  pages         = {011026},
  year          = {2017},
  doi           = {10.1103/PhysRevX.7.011026}
}

@article{Heyl2017,
  author        = {Heyl, Markus},
  title         = {Quenching a quantum critical state by the order parameter: Dynamical quantum phase transitions and quantum speed limits},
  journal       = {Phys. Rev. B},
  volume        = {95},
  pages         = {060504},
  year          = {2017},
  doi           = {10.1103/PhysRevB.95.060504}
}

@article{Zhang2017,
  author        = {Zhang, J. and Hess, P. W. and Kyprianidis, A. and Becker, P. and Lee, A. and Smith, J. and Pagano, G. and Potirniche, I.-D. and Potter, A. C. and Vishwanath, A. and Yao, N. Y. and Monroe, C.},
  title         = {Observation of a discrete time crystal},
  journal       = {Nature},
  volume        = {543},
  pages         = {217--220},
  year          = {2017},
  doi           = {10.1038/nature21413}
}

@article{Choi2017,
  author        = {Choi, Soonwon and Choi, Joonhee and Landig, Renate and Kucsko, Georg and Zhou, Hengyun and Isoya, Junichi and Jelezko, Fedor and Onoda, Shinobu and Sumiya, Hitoshi and Khemani, Vedika and {von Keyserlingk}, Curt and Yao, Norman Y. and Demler, Eugene and Lukin, Mikhail D.},
  title         = {Observation of discrete time-crystalline order in a disordered dipolar many-body system},
  journal       = {Nature},
  volume        = {543},
  pages         = {221--225},
  year          = {2017},
  doi           = {10.1038/nature21426}
}

@article{KosiorSacha2018,
  author        = {Kosior, Arkadiusz and Sacha, Krzysztof},
  title         = {Dynamical quantum phase transitions in discrete time crystals},
  journal       = {Phys. Rev. A},
  volume        = {97},
  pages         = {053621},
  year          = {2018},
  doi           = {10.1103/PhysRevA.97.053621}
}

@article{Fogarty2020,
  author        = {Fogarty, Thom{\'a}s and Deffner, Sebastian and Busch, Thomas and Campbell, Steve},
  title         = {Orthogonality catastrophe as a consequence of the quantum speed limit},
  journal       = {Phys. Rev. Lett.},
  volume        = {124},
  pages         = {110601},
  year          = {2020},
  doi           ={10.1103/PhysRevLett.124.110601}
}

@article{Sierant2023,
  author        = {Sierant, Piotr and Lewenstein, Maciej and Scardicchio, Antonello and Zakrzewski, Jakub},
  title         = {Stability of many-body localization in {Floquet} systems},
  journal       = {Phys. Rev. B},
  volume        = {107},
  pages         = {115132},
  year          = {2023},
  doi           = {10.1103/PhysRevB.107.115132}
}

@article{Iemini2024,
  author        = {Iemini, Fernando and Fazio, Rosario and Sanpera, Anna},
  title         = {{Floquet} time crystals as quantum sensors of ac fields},
  journal       = {Phys. Rev. A},
  volume        = {109},
  pages         = {L050203},
  year          = {2024},
  doi           = {10.1103/PhysRevA.109.L050203}
}

@article{Suman2024,
  author        = {Suman, M. and Aravinda, S. and Modak, Ranjan},
  title         = {Probing quantum phase transitions via quantum speed limits},
  journal       = {Phys. Rev. A},
  volume        = {110},
  pages         = {012466},
  year          = {2024},
  doi           = {10.1103/PhysRevA.110.012466}
}

@article{Zaletel2023,
  author  = {Zaletel, Michael P. and Lukin, Mikhail and Monroe, Christopher and Nayak, Chetan and Wilczek, Frank and Yao, Norman Y.},
  title   = {Colloquium: Quantum and classical discrete time crystals},
  journal = {Rev. Mod. Phys.},
  volume  = {95},
  pages   = {031001},
  year    = {2023},
  doi     = {10.1103/RevModPhys.95.031001}
}

@article{Atas2013,
  author  = {Atas, Y. Y. and Bogomolny, E. and Giraud, O. and Roux, G.},
  title   = {Distribution of the ratio of consecutive level spacings in random matrix ensembles},
  journal = {Phys. Rev. Lett.},
  volume  = {110},
  pages   = {084101},
  year    = {2013},
  doi     = {10.1103/PhysRevLett.110.084101}
}

@article{OganesyanHuse2007,
  author  = {Oganesyan, Vadim and Huse, David A.},
  title   = {Localization of interacting fermions at high temperature},
  journal = {Phys. Rev. B},
  volume  = {75},
  pages   = {155111},
  year    = {2007},
  doi     = {10.1103/PhysRevB.75.155111}
}

@article{Campostrini2014,
  author  = {Campostrini, Massimo and Pelissetto, Andrea and Vicari, Ettore},
  title   = {Finite-size scaling at quantum transitions},
  journal = {Phys. Rev. B},
  volume  = {89},
  pages   = {094516},
  year    = {2014},
  doi     = {10.1103/PhysRevB.89.094516}
}

@article{Aramthottil2021,
  author  = {Aramthottil, Adith Sai and Chanda, Titas and Sierant, Piotr and Zakrzewski, Jakub},
  title   = {Finite-size scaling analysis of the many-body localization transition in quasiperiodic spin chains},
  journal = {Phys. Rev. B},
  volume  = {104},
  pages   = {214201},
  year    = {2021},
  doi     = {10.1103/PhysRevB.104.214201}
}

@article{DeffnerLutz2013,
  author        = {Deffner, Sebastian and Lutz, Eric},
  title         = {Energy--time uncertainty relation for driven quantum systems},
  journal       = {J. Phys. A: Math. Theor.},
  volume        = {46},
  pages         = {335302},
  year          = {2013},
  doi           = {10.1088/1751-8113/46/33/335302}
}

@article{SchindlerBukov2025,
  author        = {Schindler, Paul M. and Bukov, Marin},
  title         = {Geometric {Floquet} theory},
  journal       = {Phys. Rev. X},
  volume        = {15},
  pages         = {031037},
  year          = {2025},
  doi           = {10.1103/7l91-gw77}
}

@article{Schmid2024,
  author  = {Schmid, Harald and Penner, Alexander-Georg and Yang, Kang
             and Glazman, Leonid and von Oppen, Felix},
  title   = {Robust spectral {$\pi$} pairing in the random-field
             {Floquet} quantum {Ising} model},
  journal = {Phys. Rev. Lett.},
  volume  = {132},
  pages   = {210401},
  year    = {2024},
  doi     = {10.1103/PhysRevLett.132.210401}
}

@article{Penner2025,
  author  = {Penner, Alexander-Georg and Schmid, Harald
             and Glazman, Leonid I. and von Oppen, Felix},
  title   = {Subharmonic spin correlations and spectral pairing in
             {Floquet} time crystals},
  journal = {Phys. Rev. B},
  volume  = {111},
  pages   = {184308},
  year    = {2025},
  doi     = {10.1103/PhysRevB.111.184308}
}

@article{LazaridesDasMoessner2014,
  author  = {Lazarides, Achilleas and Das, Arnab and Moessner, Roderich},
  title   = {Equilibrium states of generic quantum systems subject to periodic driving},
  journal = {Phys. Rev. E},
  volume  = {90},
  pages   = {012110},
  year    = {2014},
  doi     = {10.1103/PhysRevE.90.012110}
}

@article{AbaninDeRoeckHoHuveneers2017,
  author  = {Abanin, Dmitry A. and {De Roeck}, Wojciech and Ho, Wen Wei and Huveneers, Fran{\c c}ois},
  title   = {Effective Hamiltonians, prethermalization, and slow energy absorption in periodically driven many-body systems},
  journal = {Phys. Rev. B},
  volume  = {95},
  pages   = {014112},
  year    = {2017},
  doi     = {10.1103/PhysRevB.95.014112}
}
\end{document}